\documentclass[useAMS]{aa}

\usepackage[varg]{txfonts}
\usepackage{definitions}
\usepackage{hyperref}
\usepackage{rotating}
\usepackage{graphicx}
\usepackage{amsbsy}
\usepackage{xcolor}
\usepackage{upgreek}

\newcommand{\mr}{\mathrm}
\newcommand{\dd}{\, \mathrm{d}}
\newcommand{\bq}{\begin{equation}}
\newcommand{\eq}{\end{equation}}

\begin{document}

\title{21 cm intensity mapping with the {Five} hundred metre {Aperture} {Spherical} {Telescope}}
\author{George~F.~Smoot\inst{1,2,3}
\and Ivan Debono\inst{1}}
\institute{Paris Centre for Cosmological Physics, APC, AstroParticule et Cosmologie, Universit\'e Paris Diderot, CNRS/IN2P3, CEA/lrfu, Observatoire de Paris, Sorbonne Paris Cit\'e, 10, rue Alice Domon et L\'eonie Duquet, 75205 Paris Cedex 13, France. \\\email{ivan.debono@apc.univ-paris7.fr}
\and Physics Department and Lawrence Berkeley National Laboratory, University of California, Berkeley, CA 94720, USA. \\\email{gfsmoot@lbl.gov} 
\and Helmut and Anna Pao Sohmen Professor-at-Large, Hong Kong University of Science and Technology, Clear Water Bay, Kowloon, Hong Kong.}

\date{\today}




\abstract{
This paper describes a programme to map large-scale cosmic structures on the largest possible scales by using the Five hundred metre Aperture Spherical Telescope (FAST) to make a 21 cm (red-shifted) intensity map of the sky for the range $0.5 < z < 2.5$.  
The goal is to map to the angular and spectral resolution of FAST a large swath of the sky 
by simple drift scans with a transverse set of beams.  
This approach would be complementary to galaxy surveys and could be completed before the Square Kilometre Array (SKA) could begin a more detailed and precise effort. 
The science would be to measure the large-scale structure on the size of the baryon acoustic oscillations and larger scale,
and the results would be complementary to its contemporary observations and significant.
The survey would be uniquely sensitive to the potential very large-scale features from inflation at the Grand Unified Theory (GUT) scale and complementary to observations of the cosmic microwave background.
}


\titlerunning{21 cm intensity mapping with FAST}
\maketitle

\section{Introduction}
\label{Intro}

It has long been recognised that mapping the 21 cm hydrogen intensity is a very good way to measure the very-large-scale perturbation power spectrum \citep{Peterson:2009}, and that on scales above 8 to 10~Mpc, the hydrogen gas represents a fair and unbiased sample of the large-scale matter distribution \citep{HI:2000}.

Using the 21 cm line of hydrogen, observed over the entire sky and across a redshift range from 0 to 5, the large-scale structure of the Universe can be mapped in three dimensions. At higher redshifts, the Lyman-alpha line of hydrogen is likely to be a better mapping tool. However, 21 cm intensity mapping is still useful, since it provides a strong overlap and a more complete map.
This can be accomplished by studying a specific intensity with a resolution $\sim 5$~Mpc, 
rather than the more detailed galaxy redshift surveys. 
Intensity mapping measures the collective emission of many galaxies without detecting and studying individual galaxies.

The Five hundred metre Aperture Spherical Telescope (FAST\footnote{\href{http://fast.bao.ac.cn/en}{http://fast.bao.ac.cn/en}}) \citep{FAST_prep, Nan:2011} is a multi-beam radio telescope scheduled for completion in 2016, which can potentially be used for 21
cm surveys.

FAST is currently under construction in Quizhou province, in southwest China. It will have 4600 triangular panels and will be similar to the Arecibo Observatory, using a natural hollow for the telescope dish. The telescope will use an active surface that adjusts shape to create parabolas in different directions, with an effective dish size of 300 m. It will be capable of covering the sky within a 40-degree angle from the zenith. 
Its working frequency will be 0.3 to 3.0 GHz, with a pointing precision of 4 arcseconds (see e.g. \citealt{FASTNan2013} for technical specifications). 

FAST data will be quite complementary to galaxy redshift and weak-lensing surveys. 
The FAST data do well at covering the large scales ( $> 10$ Mpc, or $ k < 0.003\,h {\mr{Mpc}}^{-1}$), 
while the galaxy and weak-lensing surveys are more precise at small scales ($< 100$ Mpc, $k > 0.01\,h {\mr{Mpc}}^{-1}$). 
The data set can be analysed to determine baryon acoustic oscillation (BAO) features and power spectra, 
in order to address some important questions in cosmology, including dark energy, modified gravity, and inflation theory. The same data set can be used to search for and catalogue time-variable and transient radio sources.

Most importantly, studying the hydrogen emission is probably our only real tool, in addition to the cosmic microwave background (CMB), 
for studying perturbations at very large scales and to detect the features that might appear here for some inflationary potential models.

The large-scale structure of the Universe supplies crucial information about physical processes at early times. 
Unresolved maps of the intensity of 21 cm emission from neutral hydrogen HI at redshifts $z \sim 1$ to $5$ are the best hope of accessing new ultra-large-scale information, which is directly related to the early Universe.
It is possible to search for very-large-scale features induced by such processes as the GUT symmetry-breaking.
An appropriate HI intensity experiment may detect the large-scale effects of primordial non-Gaussianity, 
placing stringent bounds on different models of inflation. 
It may be possible to place tight constraints on the non-Gaussianity parameter $f_{\mr{NL}}$, with
an error close to $\sigma_{f_{\mr{NL}}} \sim 1$ \citep{Camera:2013}.

This paper describes a possible programme to map large-scale structures on the largest possible scales using FAST to make a 21 cm (red-shifted) intensity map of the sky. We focus on two anticipated science outputs: constraints on the primordial power spectrum on large scales, and BAO perturbations. The paper is organised as follows. In Sect. \ref{Goals} we outline the aims of the survey programme. In Sect. \ref{Methods} we describe the techniques and the instrument properties, including the system sensitivity, the pixel size, the drift and noise, the expected signal level, the resolution, and the foregrounds. We describe our two proposed survey strategies in Sect. \ref{Survey}. Our main results are presented in Sect. \ref{Results}. We present our concluding remarks in the last section.

 \section{Goals and outline}
 \label{Goals}
 
The possibility of performing a low-redshift spectroscopic galaxy redshift survey using FAST has been described in \citet{Duffy:2008}.
The goal we describe here is to make an intensity map of 21 cm hydrogen emission over a significant redshift range
 ($z$ from about 0.5 to 2.5, which corresponds to a frequency coverage of about 400~MHz to 1~GHz) with sufficient angular
  and frequency resolution to observe both the BAOs and any features in the very-large-scale 
  primordial perturbation spectrum.  

We show that mapping a large swath of the sky to the angular and spectral resolution of FAST by simple drift scans with a transverse set of beams can be sufficient to meet these goals. 
This paper outlines the main features necessary:  receivers with bandwidths stretching from 1.42 GHz 
(1 GHz would probably be sufficient to extract most of the scientific information as the volume at lower redshifts is not very large) 
to as low as is reasonable, but certainly to about 0.4 GHz.  This covers part of the {L~band} and ultra-high frequency (UHF) (the {L~band} is usually defined as $1-2\,\mr{GHz}$ and UHF as $0.3-1\,\mr{GHz}$). Receivers with better performance are available in the L band, with a noise temperature of around 4~K with cryogenic cooling. The main problem is loss in the feed cables and antenna structure, as well as a few additional degrees of noise from spillover and atmospheric temperature.
In the longer term, it may well work to extend the bandwidth to even lower frequencies and thus to higher redshift.

It would be best to have a system temperature of about 20K or lower so as to map most quickly and make best use of the valuable telescope time. 
The system temperature will necessarily be slightly higher at lower frequencies because of the Galactic foreground signal.
A single 40-degree swath by drift scan takes at least 48 days of observations with a linear array of receivers. In this case we assume a set of scans that adds up to 40 degrees perpendicular to the drift scan. This can be achieved with a slightly staggered linear array of 10 antennas (2 by 5) in the north-south direction so that the Earth's rotation would sweep the linear array out into a band 10 pixels wide.
If this works well, then an extension to about  two years of observations may be implemented.
This approach would complement galaxy surveys and CHIME \citep{CHIME:2013}, 
and could be completed before the Square Kilometre Array (SKA) could begin a more detailed effort. 
The science would be to measure the large-scale structure on the size of the BAOs and on larger scales. 
The survey would be uniquely sensitive to the very-large-scale features from the GUT-scale inflationary potential and complementary to the CMB observations.

\section{Concept and techniques}
\label{Methods}
The concept is to map the sky in rest-frame 21 cm wavelength (1.42 GHz and lower) with an angular resolution of about 5 arcmin 
and $\delta z$ of approximately 0.001, or 0.1 per cent  in frequency.  
This can be translated into a real-space map and power spectrum of the primordial perturbations, 
which we believe are from inflation (e.g. \citealt{Liddle:1999}). 
These perturbations are processed through the BAOs and the damping of structure formation. 
We believe that we understand these processes, but the data can be used first to check the BAOs, 
using them as a standard ruler and testing the dark energy acceleration of the Universe and potential modifications to General Relativity.  
If these are in agreement, the primordial power perturbations can easily be unfolded and tested against the predictions of inflation and GUT-perturbed inflationary models as the features should show up both in real space and in the power spectrum.

 \begin{table*}
\caption{Main survey and instrument specifications. Most of these details are provided in \citet{Nan:2011} and \citet{FASTNan2013}. Wherever our calculations assume some figures that are slightly different from the official FAST specifications, these are given in square brackets. We note that the project anticipates that more than 19 receivers and antennas will be built. The main factor that will determine this is the cost and effort of building and installing them. }
\label{Specs}
\begin{center}
\begin{tabular}{ll}
\hline
\hline\\
Radius of reflector & 300~m \\
Aperture of dish &  500~m \\
Illuminated aperture & 300~m [500~m]\\
Opening angle & 100 degrees to 120 degrees\\
Zenith angle & 40 degrees\\
Tracking range & 4 to 6 hours\\
Frequency coverage & 0.4 to 1.42~GHz\\
Total system temperature ($T_{\mr{sys}}$) &20 K\\
Sensitivity (L band) & Antenna effective area/system noise temperature $\sim 2000\,\mr{m}^{2}\mr{K}^{-1}$ \\
Angular resolution (L band) & 2.9 arcmin [we assume 5 arcmin over the range 0.4 to 1.42~GHz]\\
Number of beams (L band) & 19 [we assume more than 19 receivers and feeds] \\
Pointing accuracy & 8 arcsec\\
Integration time per pixel & 4 seconds\\
Pixel bandwidth in $z$ direction & 0.3 per cent\\
Redshift range of survey & 0.5 to 2.5\\
\hline
\end{tabular}
\end{center}
\end{table*}

\subsection{Required system sensitivity}

The relative brightness from the major background of synchrotron radiation is estimated to be about 0.3 to 1~K at 1.4~GHz near the Galactic plane, down to a substantially lower value as we move well off the plane (see e.g. \citealt{SETI:2007} for detailed figures). 
At 0.41~GHz it is about 3 to 10~K. This means that the total background temperature including CMB is about 4 to 15~K 
and that the receivers should not contribute significantly to this total, or it would slow the mapping speed of the 
telescope.
The aim should be for a complete system with an effective noise temperature of about 20~K.

\subsection{Pixel size, volume, and sensitivity}
We consider a pixel size of 5~$h^{-1}$~Mpc as our standard-sized comoving sample.
(This gives a Nyquist-size sample for 10~$h^{-1}$~Mpc wavelengths and above.)
Assuming an integration time $t_\mr{pix}$ of 4 seconds per pixel, the Earth's rotation will have swept through about 1~arcmin during this time interval. In 20 seconds the rotation of the earth sweeps through our canonical 5~arcmin, which in the redshift range of 1 to 2 typically gives 
the 5~$h^{-1}$~Mpc size. Thus a single drift scan integration time is about 20 seconds.

In the redshift direction, a 5~$h^{-1}$~Mpc in depth pixel has an error of
$\delta z = 5 h^{-1}\mr{Mpc}  \, H /c \sim 0.001  $, or $0.1$ per cent. This gives us our pixel bandwidth $\Delta f$. Here, and throughout the text, $h$ is the dimensionless Hubble parameter, $H$ is the Hubble constant such that $H=100h\,\mathrm{km\,s^{-1}Mpc^{-1}}$, and $c$ is the speed of light.

The 21 cm hydrogen line from a source at redshift $z$ has a frequency of $1.42\,\mr{GHz}/(1+z)$. Using these figures, and assuming a total system temperature of 20~K, we have for each receiver channel
\begin{align} 
\delta T_{\mr{rms} ~\mr{pixel}} & =\frac{T_{\mr{sys}}}{\sqrt{\Delta f t_{\mr{pix}}}}  \nonumber \\
&=\frac{20\,\mr{K} }{ (0.001 \times (1.42\times10^{9} / {(1+z)})\times 20)^{1/2}} \nonumber \\
&=  3.75 {(1 + z)}^{1/2}~\mr{mK}\, .
\end{align}
The BAO scale is around 100~Mpc. This means that we obtain a reasonable number of samples (about 400, and greater when the $z$ 
direction is included) in each BAO feature and thus an expected noise of around 0.2~mK for a BAO feature.
Even with the array of 20 receivers, we still lack by a factor of two in what we would desire for sensitivity.
To gain this would require at least four days per elevation scan.

\subsection{Drift and 1/f noise}
The discussion thus far has only considered receiver white noise.
In practice, for long periods there are drifts in the receiver performance, particularly in gain, that must be considered.
This is generally known as 1/f noise or flicker noise since it is the dependence that does not integrate down and can leave long-wave signals in the observations. 
The standard technique is to make a differential observation or a correlation receiver, which again means differencing two portions of the sky or an equivalent temperature reference load. 
Ideally, something like the north celestial pole region would
be used as the reference point (this is always visible from FAST), either in a difference (e.g. Dicke radiometer) or correlation receiver (see \citealt{Battye:2013}).
The Dicke radiometer causes a loss of a factor of two in sensitivity.
With the correlation receiver we can break even or lose $\sqrt{2}$ in sensitivity.

Depending upon the receiver performance, we can use either one reference antenna pointing towards the north pole, or a very broad beam pointing vertically (slowly changing with the drift scanning and thus being a local rather than global switching). When the receivers are stable enough, or when a good-quality 5 by 5 array is used, the Dicke switching and thus sensitivity loss could be avoided. Digital processing of the data and array could then take out the 1/f noise. 

A wide beam pointing directly upward might be considered to compare with the high-resolution beam through the downward-pointing feed.
This would give an average over the HI 21 cm emission that will be approximately the same over all the sky, and it would average out most of the extragalactic point sources. It also looks through the same column of atmosphere, allowing cancellation of that signal.
However, the Galactic synchrotron emission varies on the very large scale on the sky so that there would be primarily a large quadrupolar feature in the observations plus an additional signal on the Galactic plane. 
This can probably be handled to first order by a smooth parameterisation modelling.

Frequency switching and/or using symmetric matching beams can
be considered as references, but this requires a somewhat more complicated scanning and data processing system. However, it may well be worth the effort if the differential system over-complicates the receiver cabin.  It would be good to have the knee or corner frequency below the switching frequency so that the additional noise is avoided. Without switching, the 1/f knee should be at the atmospheric knee to avoid adding noise.

The opposite direction differential correlation receiver system is probably the best option, but a switching differential will also work well.
For the purposes of this paper we assume that a simple solution exists in this form and use the numbers to provide a scale and reference for a more detailed design.

Significant advances in room-temperature low-cost receivers with adequate noise figure have been made.  
In 2014 they cost as little as 60 yuan each. There have also been advances in FPGA that 
make processing the signal into 0.1 per cent  frequency bands (roughly 1 MHz) readily achievable, 
so that adding about 20 receivers covering this frequency range to the existing system can easily be envisaged.
The main cost is the effort of assembling and installing the receivers and antennas.

\subsection{Expected signal level}
There have been a number of estimates of the expected 21 cm intensity mapping signal level.
A recent paper \citep{Bull:2014} has reported the effective HI brightness temperature $T_b$  split into a homogeneous and fluctuating part, such that $T_b = \overline{T}_b (1 + \delta_\mathrm{HI})$ and 
\begin{equation}
\overline{T}_b = \frac{3}{32 \pi} \frac{h c^3 A_{10}}{k_B m_p^2 \nu_{21}} \frac{(1+z)^2}{H(z)} \Omega_\mathrm{HI}(z) \rho_{c,0}.
\end{equation}
The fluctuations are given by
\begin{equation}
\delta T^S ({\boldsymbol{\theta}}_p,\nu_p) = \overline{T}_b(z) \delta_{\mr{HI}}(\boldsymbol{\mathrm{r}}_p, z)\, , 
\end{equation}
where ${\boldsymbol \theta}_p$ is the two-dimensional direction of the signal, $\nu_p$ is the frequency, and $\boldsymbol{\mathrm{r}}_{p}$ is the location in comoving space.

Here and throughout the text, $\Omega_\mathrm{HI}$ is the HI density fraction, equal to $\rho_{\mathrm{HI}}/\rho_{c,0}$, where $\rho_{c,0}=3H^2/8\pi G$ is the critical density today. $A_{10}$ is the Einstein coefficient for spontaneous emission, $m_{p}$ is the proton mass, $\nu_{21}$ is the frequency of the 21 cm emission, $k_{B}$ is the Boltzmann constant, and $H(z)$ is the Hubble rate at redshift $z$.

The HI density contrast is
 \begin{equation}
 \delta_{\mr{HI}} = b_{\mr{HI}} \star \delta_M \, ,
\end{equation}
where $\delta_{\mr{M}}$ is the total matter density perturbation (cold dark matter and baryons), $b_{\mr{HI}} $ is the bias, and $\star$ denotes convolution.
The current best estimate of the biased neutral hydrogen density (at the $68$ per cent confidence level at redshift $0.8$) is
\begin{equation}
\Omega_{\mr{HI}}b_{\mr{HI}} = 4.3\pm{1.1} \times 10^{-4}\, . 
\end{equation}
Assuming that the peculiar velocity gradient and $v/c$ terms are small for these large pixels, we finally obtain
\begin{align}
T_{b}(\nu,\Delta\Omega,\Delta\nu) &\approx \overline{T}_{b}(z) \left(1+b_{\mr{HI}}\delta_m(z)-\frac{1}{H(z)}\frac{\dd v}{\dd s}\right) \nonumber\\
\overline{T}_{b}(z) &\approx 566h\left(\frac{H_0}{H(z)}\right)\left(\frac{\Omega_{\mr{HI}}(z)}{0.003}\right)(1+z)^2{\upmu\mr{K}}\, . \nonumber\\
\end{align}
That is to say, the typical scale of the HI 21 cm signal is around 0.5~mK.

Here we used a signal model where we averaged over a radial bin and measured the power spectrum in redshift. This is described in \citet{Battye:2012} and \citet{Battye:2013}. The projected error $\sigma_P$ on a power spectrum measurement average $P$ over a radial bin in $k$-space of width $\Delta k$ is 
\begin{equation}
\frac{\sigma_P}{P}=\sqrt{2\frac{(2\pi)^3}{V_{\mr{sur}}}\frac{1}{ 4\pi k^2\Delta k}}\left(1+\frac{\sigma_{\mr{pix}}^2 V_{\mr{pix}}}{ [\overline{T}(z)]^2W(k)^2P}\right)\,,
\label{eqn:error}\end{equation}
where $V_{\mr{sur}}$ is the survey volume, $V_{\mr{pix}}$ is the pixel volume, $\sigma_{\mr{pix}}$ is the pixel noise over a nominal bandwidth of $\Delta f=1\mr{MHz,}$ and $W(k)$ is the angular window function. The latter is given by
\begin{equation}
W(k)=\mr{exp}\left[-\frac{1}{2}k^{2}r(z)^{2}\left( \frac{\theta_{\mr{FWHM}}}{\sqrt{8\ln 2}}\right)^{2} \right]\, ,
\end{equation} 
where we assumed an angular resolution $\theta_{\mr{FWHM}}$ of 5~arcmin. 
The survey volume is a function of the observed patch of the sky $\Omega_{\mr{sur}}$ and is given by
\bq
V_{\mr{sur}}=\Omega_{\mr{sur}}\int_{z_{\mr{min}}}^{z_{\mr{max}}} \dd z \frac{\dd V}{\dd z \dd \Omega}\,,
\eq
where \bq
\frac{\dd V}{\dd z \dd \Omega}=\frac{c r(z)^2 }{H_0E(z)}\,.
\eq
The average temperature is 
\begin{align}
\overline{T}(z)&=44\,\upmu{\mr{K}}\left(\frac{\Omega_{\mr{HI}}(z) h}{2.45\times 10^{-4}}\right)\frac{(1+z)^2}{E(z)}\nonumber\\
&= 440\,\upmu{\mr{K}} \left(\frac{H_0}{H(z)}\right) \left(\frac{\Omega_{\mr{HI}}(z) h}{ 2.45\times 10^{-3}}\right){(1+z)^2}\, ,
\end{align}
where $\Omega_{\mr{HI}}$ is the HI density relative to the present day critical density, and $E(z)=H(z)/H_0$.

The first part of Eq. \ref{eqn:error} is the sample variance of the number of modes included in the survey. 
The second term takes into account the signal-to-noise ratio on a pixel.
That is to say, if the noise variance on a pixel is larger than the signal in the pixel, 
the net error on the power spectrum is increased by that factor. 
This is the reason why subduing the receiver noise down to  a minimum addition to the background is so relevant.

\subsection{Mapping angular resolution to spatial resolution}

We considered the angular resolution of FAST.
For a full dish illumination, the resolution at 21 cm wavelength is around
\begin{align}
\theta &=1.22 \lambda / d = 1.22(0.21 / 500)\nonumber\\& = 5.21 \times 10^{-4}\,\mr{radians} = 1.76\,\mr{arcmin}\,.
\end{align}
The angular resolution scales with redshift as $\theta(z) =1.2 \lambda(z) / d  = 1.76 (1+z)$~arcmin.  
Thus, the angular resolution increases from 1.76~arcmin to just over 5~arcmin at a redshift of 2, corresponding to a wavelength of 
$\lambda(z) = (1 + z)$ 21 cm, or 
$\lambda_{z=2}= 63$~cm corresponding 
to a frequency $f = 1.42\,\mr{GHz} / (1 + z) $, or $ f_{z=2} = 473$~MHz.
For $z = 2.5$, these values are $ f_{z=2.5} = 405.7$ MHz and $\lambda_{z=2.5}= 73.5$~cm,
and for $z = 0.5$, $ f_{z=0.5} = 946.7$ MHz and $\lambda_{z=0.5}= 31.5$~cm.
Now it is possible to use the full 500 m aperture for the vertical drift scans. To extend the width of the survey,
it is necessary to accept a lower illumination away from
the centre and probably to build 
the proposed ground screen especially on the north side of the telescope to look farther south.

The angular resolution is related to the physical size of observed objects.
In considering the relation between angular size and cosmological redshift for the purpose of our survey, we assumed the current $\Lambda$CDM concordance model of the Universe: flat, with a cosmological constant and cold dark matter content such that $\Omega_{\mr{m}} +  \Omega_{\Lambda} = 1$. For the current best-fit value of $\Omega_{\mr{m}} \sim 0.3$, the angular size of objects in the range 
  $1 < z <      3$ is relatively constant (see e.g. \citealt{Sahni:2000}). For instance, a galaxy 
cluster with a size of about 1~Mpc will never subtend an angle less than about $4\, h$~arcmin, regardless of its
 distance from the observer. Thus FAST could also be used to search for galaxy clusters in a way that is complementary 
 to the SZ searches, with the advantage that mapping the intensity is a more direct and straightforward way to observe the large-scale 
 structure.

The BAO scales corresponds to angles of about 2$\degr$ at $z = 2.5$ and 4$\degr$ at $z = 0.8$ 
(see Table \ref{tab:angle} for some specific values).
Along the line of sight, at redshift 0.8 (800 MHz) the BAO scale corresponds to a correlation at 20~MHz separation, 
and at redshift 2.5 (400 MHz),  it corresponds to a 12~MHz separation correlation.
Here we anticipate  $\sim 5$~arcmin, or more specifically, ${5\,h^{-1}\mr{Mpc}}$
sized pixels as a standard and thus have very many samples.

The BAO scale is $\sim 100$~Mpc. We need to sample it well to obtain science results at the 1 per cent order of magnitude. This gives us a critical size of around 10~Mpc.
At a redshift of $z = 2$ the size 10 Mpc is about 19 arcmin in angle or slightly larger.
Thus we have a 5 arcmin resolution.
This gives us sampling slightly better than Nyquist, which is good for making maps and power spectra.

\begin{table*}
\caption{Angular size for two critical comoving sizes:
10 ${h^{-1}\mr{Mpc}}$ and the BAO standard ruler of 150 ${h^{-1}\mr{Mpc}}$. 
Since they are expanding, the angle does not stay constant as it does for the simple fixed physical distance of 10 Mpc. 
The comoving angle reaches relatively constant value at higher redshift. The cosmology assumed here is a flat universe with
$\Omega_{\mr{m}} = 0.3036$ and $H_0 = 68.34\,\mr{km}s^{-1}\mr{Mpc}^{-1}$.} 
\label{tab:angle} 
\centering
\begin{tabular}{l l l l }
\hline\hline
 Redshift $z$ & Angle 10 Mpc   &   Angle 10 $h^{-1}\mr{Mpc}$  & Angle 150 $h^{-1}\,\mr{Mpc}$\\
                        &  (arcmin) &   (arcmin)  & (degree) \\
\hline
0.5 &26.56  & 26  & 6.5  \\
1 & 20.3  &14.9  & 3.72  \\
2 & 19.2  & 9.5  & 2.38  \\
3 & 21.1  & 7.75  & 1.94  \\
4 & 23.4  & 6.87  & 1.72  \\
5 & 25.9  & 6.34  & 1.59  \\
\hline
\end{tabular}
\end{table*}

\subsection{Foregrounds}

There are a number of foregrounds, and of possible values for these foregrounds. We discuss the CMB, the atmospheric background, Galactic synchrotron  and free-fall thermal emission, and extragalactic radio sources. In our survey calculations, we only include that part of the sky where Galactic foregrounds are minimal.

The CMB has a flat antenna temperature of about 2.72~K for the frequencies of interest here.
This is not a concern here except for the added noise and the need for the HI emitters to be at a different temperature.

The atmospheric background has, at 1.42~GHz, a typical antenna temperature of about 2~K looking to the zenith.
The signal generally decreases as the frequency squared and is thus less important for $z = 1$ by a factor of 4.
The problem is the small variation and factors such as precipitation and mist or dew.  

The main foreground is Galactic synchrotron emission. There is also some foreground on the Galactic plane, which is mostly HII thermal emission.
We can approximate the signal from the Galactic synchrotron as
$T_{A ~\mr{synchrotron}} \sim (1 \text{\,to\,} 6) \times  (\nu/ 1 ~\mr{GHz})^{-2.7}\,\mr{K}$; the range reaches from the north Galactic pole down to 10 degrees within the plane of the galaxy. 

We can convert this into thinking in redshift $T_{A ~\mr{synchrotron}} \sim (1 \text{\,to\,} 6) \times  (1 +z )^{2.7} $ K.
Thus, at very low redshifts the Galactic synchrotron is roughly 10$^3$ to $10^{4}$ brighter than the strongest signal we expect and 
10$^4$ to 10$^5$ times the signal level we hope to observe.
At a redshift of $z = 1$, that would be 5 times worse, and for $z = 2$, it would be 16 times worse.
Thus we have to think about how to model and remove this foreground.

Galactic synchrotron emission is radiation produced by high-energy cosmic-ray electrons above a few MeV 
spiralling in the Galactic magnetic field (e.g. \citealt{Pacholczyk:1970,Banday:1990,Banday:1991}). 
There is lower emission in relatively smooth regions away from the Galactic plane and Galactic loops. 
Based upon the synchrotron mechanism, we anticipate a running power-law in frequency for the Galactic synchrotron brightness temperature:
\begin{equation}
T_{\mr{syn}}=A_{\mr{syn}}\left(\frac{\nu}{\nu_\ast}\right)^{-\beta_{\mr{syn}}-\Delta\beta_{\mr{syn}}\log(\nu/\nu_\ast)},
\label{eq:Tsyn}
\end{equation}
where $A_{\mr{syn}}$ is synchrotron brightness temperature at $\nu_\ast=1$ GHz, and $\beta_{\mr{syn}}$ and $\Delta \beta_{\mr{syn}}$
are the spectral index and spectral running index, respectively. 
From the 408 MHz all-sky continuum survey of \citet{Haslam:1981,Haslam:1982}, \citet{Haverkorn:2003} estimated the mean brightness temperature at 408 MHz to be $\sim 33$~K with temperature uncertainty of $\sim 10$ per cent. 
There is some uncertainty in the zero point of the map.
After subtraction of the $\sim 2.7$~K contribution of the CMB and the $\sim 3.1$~K contribution of extragalactic sources (e.g. \citealt{Bridle:1967, Lawson:1987, Reich:1988}), 
the diffuse synchrotron Galactic background is $\sim 27.2$~K at 408~MHz. 
For a spectral index of 2.74 \citep{Platania:2003} the estimated synchrotron brightness temperature 
at 1 GHz would be $A_{\mr{syn}}=2.4\pm 0.24$~K. 
At high Galactic latitudes, the brightness temperature has a minimum of $\sim 1$~K \citep{Lawson:1987,Reich:1988,Shaver:1999}. 
A wider range of estimates for the mean spectral index  reaches from 2.6 to 2.8 
(e.g. \citealt{Bridle:1967,Willis:1977,Lawson:1987,Reich:1988,Banday:1990,Banday:1991, Tegmark:2000,Platania:2003}),
with indications for dispersion at each position on the sky that
is due
to distinct components along the line of sight (e.g. \citealt{Lawson:1987,Reich:1988,Banday:1990,Banday:1991,Shaver:1999}). 

For estimation purposes, we chose a typical spectral index of $\beta_{\mr{syn}}=2.7$ with dispersion of 0.1 \citep{Reich:1988,Shaver:1999}, and spectral running index of $\Delta\beta_{\mr{syn}}=0.1$ \citep{Tegmark:2000,Wang:2006}.
\citet{Kogut:2012} has analysed the synchrotron spectral index and fitted it with an index
$\beta_{\mr{syn}} = 2.6$ and $\Delta\beta_{\mr{syn}}=0.06$.
There is clearly an allowed range, so in practice we use a number in the 2.6 to 2.7 range with some sort of weighted prior.

Free-free thermal emission comes from ionised regions in the interstellar medium (ISM), 
with electron temperatures of $T_e > 8000$~K.
Again, the physical mechanism for free-free emission implies a good approximation is a running power-law in frequency for the Galactic free-free brightness temperature \citep{Wang:2006},
\begin{equation}
T_{\mr{ff}}=A_{\mr{ff}}\left(\frac{\nu}{\nu_\ast}\right)^{-\beta_{\mr{ff}}-\Delta\beta_{\mr{ff}}\log(\nu/\nu_\ast)},
\label{eq:Tff}
\end{equation}
where $A_{\mr{ff}}$ is the free-free brightness temperature at
$\nu_\ast=1$ GHz, and $\beta_{\mr{ff}}$ and $\Delta\beta_{\mr{ff}}$
are the spectral index and spectral running index, respectively. 
We can estimate $A_{\mr{ff}}=0.12 \pm 0.01$~K assuming a 10 per cent temperature uncertainty. 
At high frequencies ($\nu>10$ GHz) the brightness temperature spectral index is
$\beta_{\mr{ff}}=2.15$, while at low frequencies it drops to
$\beta_{\mr{ff}}=-2.0$ because of optically thick self-absorption \citep{Bennett:2003}. 
Since the gas is optically thin above a few MHz, the
brightness temperature spectrum  is
well described with a spectral index of $\beta_{\mr{ff}}=2.10\pm 0.01$
\citep{Shaver:1999}, and the spectral running index is
$\Delta\beta_{\mr{ff}}=0.01$ \citep{Tegmark:2000,Wang:2006}.

Regarding extragalactic radio sources, the survey should use existing maps of radio sources and its own ability to distinguish 
point sources to block out those pixels containing them \citep[see e.g.][]{ARCADE2skybrightness,ARCADE2skybrightness2}.
The problem of the confusion limit can be handled by the large beam size and power spectrum 
treatment and exclusion. 
We can discard the large signal sources, and there are many radio surveys
with much higher angular resolution.
Their confusion limit will be averaged over by the larger FAST beam size. We can make use of their 
frequency and spatial power spectrum to estimate and remove their effects.

The total emission of extragalactic foregrounds has been estimated both directly and from integrated source counts. 
At 150 MHz, its contribution to the contamination varies from $\sim 30$~K \citep{Willis:1977,Cane:1979} 
to $\sim 50$~K \citep{Bridle:1967,Lawson:1987,Reich:1988}. 
At 1 GHz (midway between $z = 0$ and $z = 1$), the estimate is roughly 0.15 K.
These foregrounds produce $\sim 10$ per cent of the total contamination on average, 
but can reach $\sim 25$ per cent at high Galactic latitudes, 
at the minimum brightness temperature of the diffuse Galactic emission.

Most foregrounds are due to radio point sources and are related to active galactic nuclei activity.  
Radio halos and radio relics are also significant foregrounds, 
but since they appear only in rich galaxy clusters, they are rare.

\subsection{Removal of foregrounds}
There is an extensive literature on the removal of foregrounds from 21 cm intensity mapping (e.g. \citealt{Battye:2013,Alonso:2014,Bull:2014, Shaw:2014}).
We argue here that the real science gain in general, and for FAST in particular, 
is 21 cm intensity mapping in the range $0.5 < z <2.5$, 
where the foreground emission is significantly lower and easier to understand.

The usual discussion of how to treat the diffuse Galactic foregrounds is 
that the synchrotron sources, radio sources, and HII emissions are very smooth in frequency.  
The 21 cm signal we seek and which we wish to observe varies rapidly with frequency. 
Thus we fit a smooth background composed of 
\begin{align}
T_{\mr{foreground}} &= T_{\mr{CMB}} + T_{\mr{atm}} \nu^2 + T_{\mr{syn}} \nu^{-(2.7+ \delta \beta_{\mr{syn}})} +\nonumber \\ &T_{\mr{ff}} \nu^{-2.1} 
+ T_{\mr{exgal}} \nu^{-2.7}
+ {\mr{baseline~polynomial}} .
\end{align}
Each of these can be fitted with a slightly varying index as we are not concerned with dividing the background into components as much as separating it from our desired signal.
We can make a frequency dependence fit to each stack of pixels at a given direction on the sky 
and then produce a cleaned pixel stack as we fit over the redshift direction.
We would then remove the baseline and obtain a differential temperature map of the anticipated 21 cm signal.
In the fitting we can add the priors on the expected level of signal and probably just include the extragalactic signal as a portion of the synchrotron and the free-free emission, which reduces the number of parameters for which to fit.

Another approach is to carry out the 3D Fourier transform, and collapse the transverse directions on the sky and plot $k_\perp$ versus
$k_\parallel$ amplitudes and then stamp out the galaxy as being all at very low $k$.

The challenge in the foreground removal is that we may lose some of the very largest-scale power in the $z$-direction.
There are typically around 700 data points in the redshift direction, and thus one has 700 pixel values plus about eight parameters to determine, which leaves us with some freedom and a reduction in power of the desired signal on the scales of the allowed degrees of freedom. 
This can be improved upon by a combined fit in three dimensions and allowing only the 21 cm signal and the extragalactic sources to vary on the small scales and requiring the other sources,
such as CMB, atmosphere, synchrotron, and free-free emission,
to vary only slowly in the transverse directions. 
For the atmosphere a monitor or atmospheric emission estimation system is required.
The correlation system with larger angle vertical antenna can provide information on the timescales of the correlations, the frequency dependence, and the spatial dependence
if so exploited, but correlations in the receiver array complex and other indicators may also be used. Likewise, regular calibrations or scans across various sources can provide information on the allowed baseline shapes and changes with time or position.

In general, there are the signals that do not change much with frequency, including the CMB and the receiver noise,
and those with a fairly strong frequency dependence.
These signal totals run on the scale of 30 K, and we aim to achieve the 1 mK level and thus need to make the subtraction or removal at the $10^{-4}$ level or better.
This might be possible by examining the power spectrum in angle and particularly in frequency.

The foregrounds must be removed to one part in $10^4$ to be below the thermal noise in the maps
and to achieve the primary science goals.
This will require a careful determination of the absolute calibration and of the beams as a smooth function of frequency.

Some of the important challenges (including side lobes, calibration, and cross-polarisation) are specific to the instrument. A precise estimate of these effects requires detailed engineering specifications, as well as observation strategy and conditions (see \citealt{Dong2013a,Dong2013b}).  As this information becomes available, accurate simulations of the instrument performance will become possible.

\section{Survey strategy}
\label{Survey}

For the purpose of this paper, we consider two indicative surveys. The first is a proof-of-concept 2D map. The second is a large-scale 3D perturbation map, for which we consider three variations. These are described below. Table \ref{tab:surveys} provides a list of the surveys referred to in this work, together with their survey configuration.

\begin{table*}
\caption{Indicative FAST surveys considered in this work.} 
\label{tab:surveys}
\centering
\begin{tabular}{l l l l l l }
\hline \hline
 Name & Type & Observing time & Declination strip width & Observed sky area & Survey volume \\ 
 &&&(degrees) & (sq. degrees) &(Gpc$^{3}$)\\
 \hline
1 & 2-D  & 2 weeks & $0.5\degr$&45 &$0.3$\\
2a & 3-D  & 50 days & $40\degr$ &3600 &$24$\\
2b & 3-D & 2 years & $40\degr$ &3600 &$24$\\
2c & 3-D & 4 years &  $80\degr$ &7200 &$48$\\
\hline
\end{tabular}
\end{table*}

\subsection{Survey 1: proof-of-concept 2D map}
 We first consider the result of a continuous drift scan. 
That is, the scan is performed by pointing the FAST beams close to vertical and letting the Earth's rotation sweep the beam across the sky at a rate of 15$\degr$ per hour. The resulting survey volume has the shape of a very thin wedge (just 0.5 degrees in width), with one dimension given by the angle swept by the rotation of the Earth, and the one dimension given by the redshift. Since the volume is almost a plane, we refer to it as a 2D map.
The observations made and the data are binned by angle on the sky and by frequency.
We assume that we bin on the scale of $5\,h^{-1}$ Mpc comoving scale in angle and 0.1 per cent in frequency, giving roughly the same smooth size pixels in both dimensions. 
After a few days of scanning, how would the slice of data appear?

\citet{Peterson:2009} carried out a simulation of the fluctuations in the brightness temperature of 21 cm emission from galaxies in a slice of the Universe, in which the emission is smoothed over an $8\,h^{-1}$~Mpc scale. The redshift $z$ and frequency of observation are related by {$f = 1.42\,\mr{GHz} / (1 + z)$}. This simulation assumes a simple power law of perturbations with random phase. This is the type of map that might be made with a set of drift scans looking at the zenith or near it. 
With the suggested receiver set for FAST, we expect to make such a 2D map with a colour depth of 1 bit (5 to 10 pixels),
to a sensitivity or signal-to-noise ratio of 1 to 2 with about two weeks of scanning and serious effort at data processing and analysis to remove the foreground contamination.
Such a map would provide a strong proof of concept and produce science results on the BAO scale and larger.

The map would be larger than the one shown in \citet{Peterson:2009}, extending to 90$\degr$ in angle and for $0.5 < z < 2.5$. 
The ultimate map would depend upon how stable the system is,
however, and if data can be taken during daylight hours as well as at night.

We refer to this survey as survey 1. Even with this simple but deep survey, there are significant scientific results. It could measure both the large-scale perturbation spectrum and the BAO peaks fairly accurately
over an extended redshift range.

\subsection{Survey 2: large-scale 3D perturbation mapping}
It is straightforward but time-consuming to move to 3D perturbation mapping. The angle to the zenith is changed in the north-south direction, and the Earth's rotation provides another slice. This is achieved in FAST by moving the feeds and distorting the shape of the main dish to a parabola pointing in the direction set by the location of the feeds (see \citealt{FASTHu2013}, and \citealt{FASTJiang2015} for details of the engineering). The result is a thick wedge whose width is given by the variation in this pointing angle. The other two dimensions are given by the rotation of the Earth and the redshift. We therefore refer to this as a 3D map.

The process is substantially sped up for a north-south linear array or receivers in the receiver cabin that provides several sections per daily rotation. 
For reasons of space and preventing cross-talk, the receivers'  feeds are probably offset so that we would have to interleave at least a pair of scans to obtain a completed swath.
If the linear array were as large as 10 receivers in the north-south direction, 
then the aberration would not be too severe for the $\pm 5 \times 5 = \pm 25$~arcmin
even though FAST is a very fast telescope with an f-number of about 0.5.
To obtain a swath 40$\degr$ wide would take about N = 40$\degr$/($n$ 5~arcmin) = 48 (10/$n$)~days, 
where $n$ is the number of receivers in the north-south direction.

We refer to this survey as survey 2.

Again, we would aim to provide a detailed concept layout to determine whether a two-by-ten array or something like a five-by-five array 
would produce better performance for FAST, since its surface is actively deformed to produce a parabola for the direction of observation. Here, for both surveys 1 and 2, we assumed ten receivers. 

\section{Anticipated science}
\label{Results}

Most of the cosmology science output in intensity mapping comes from measuring the perturbations, in the map and power spectrum, and comparing 
them to theoretical predictions and expectations, or finding the parameters of the models.
The precision to which the real space map can be made and the perturbation power spectrum
can be determined decide the reach of the experimental observations.

\begin{figure*}
\centering 
\includegraphics[angle=270,width=12cm]{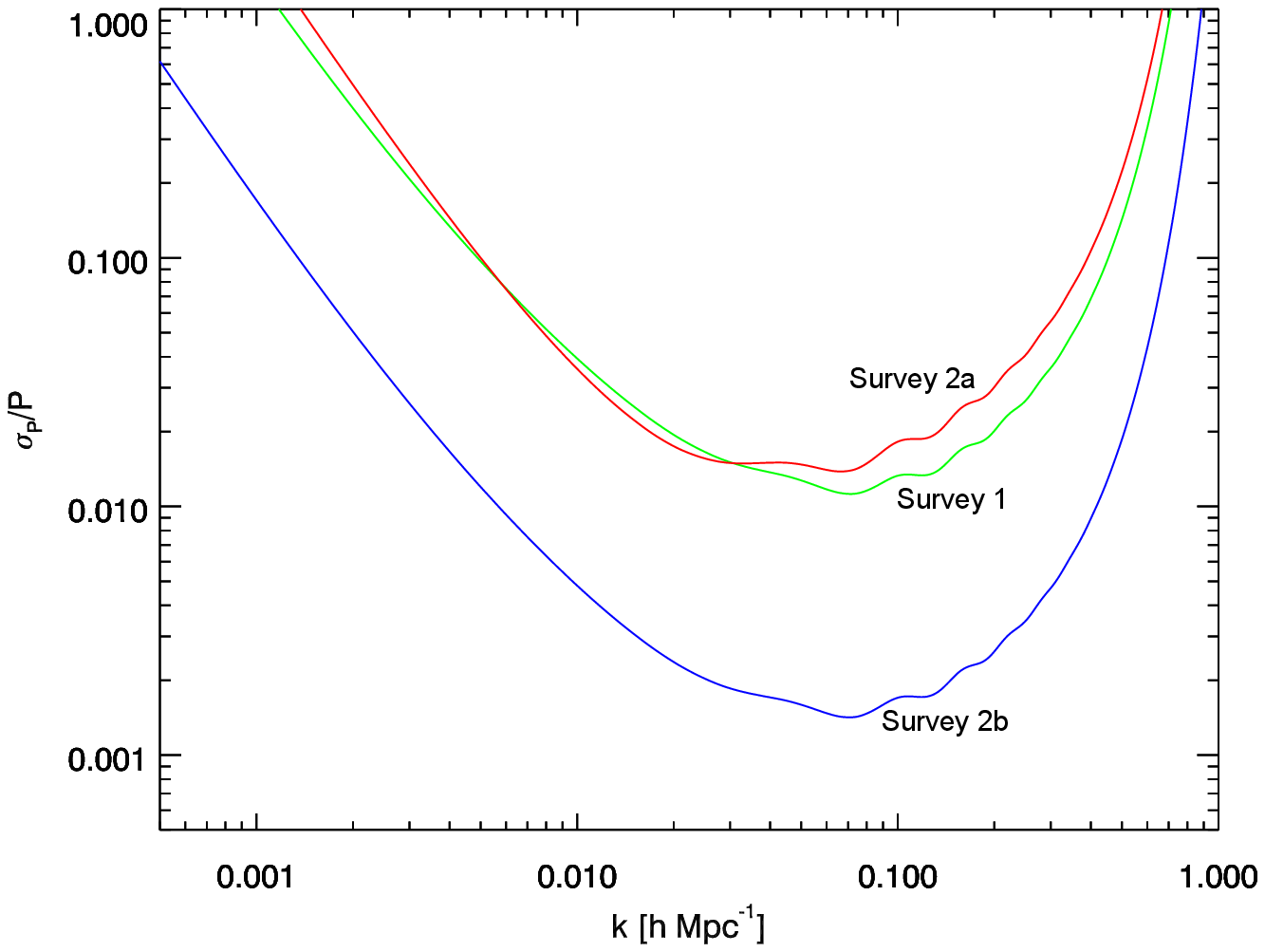}
\caption{Fractional error in the power spectrum versus wave number $k$ at a redshift of $z = 0.9$. Here we use a binning scheme with $\Delta k=1\,h{\mr{Mpc}}^{-1}$.
The two upper curves are for parameters from surveys 1 and 2, with only a first pass for survey 2. The lower curve is for survey 2, with an observing time of two years. The $z$-range is 0.5 to 2.5. The survey and pixel volume assume $\Delta z \approx 0.3$, which corresponds to our nominal frequency bandwidth of 1MHz.}
\label{fig:surveys}
\end{figure*}

Figure \ref{fig:surveys} shows the fractional errors for the two surveys versus wave number $k$.
The two plots are for parameters of surveys 1 and 2. 

Since the observations are sensitivity-limited, the errors decrease linearly with inverse of the observing time, 
but only with the inverse square root of the survey volume, and the first cut of each gives about the same errors. 
Since for short observing times the noise variance dominates, 
the much longer exposure time of survey 1 therefore makes up for the longer total exposure (but 14 times shorter per pixel) spread over a much larger volume and thus number of pixels. 
The 80 times larger volume does not quite make up for the 14 times shorter integration time per pixel.
Survey 1 was chosen to reach about unity in the second term added to one at the power spectrum peak, 
so that it can only improve marginally there with more observing time. It would improve by 25 per cent for about double the proposed two-week run, compared to the more rapid return to 14 days, and it still improves away from the power spectrum peak.
Survey 2, on the other hand, will reduce its fractional error with the inverse of the observing time in units of the base observing time.
If survey 2 is going well and integrating down properly, then it therefore reaches diminishing returns in about 14 times the original mapping time.  The lower curve shows the fractional error that would be achievable with two years of observations in survey 2 mode assuming a duty cycle of $> 80$ per cent.  
We also need to allow for the case that the effective system temperature is higher than 20 K or different scan and differencing or foreground removal strategy.

\begin{figure*}
\centering
\includegraphics[angle=0,width=12cm]{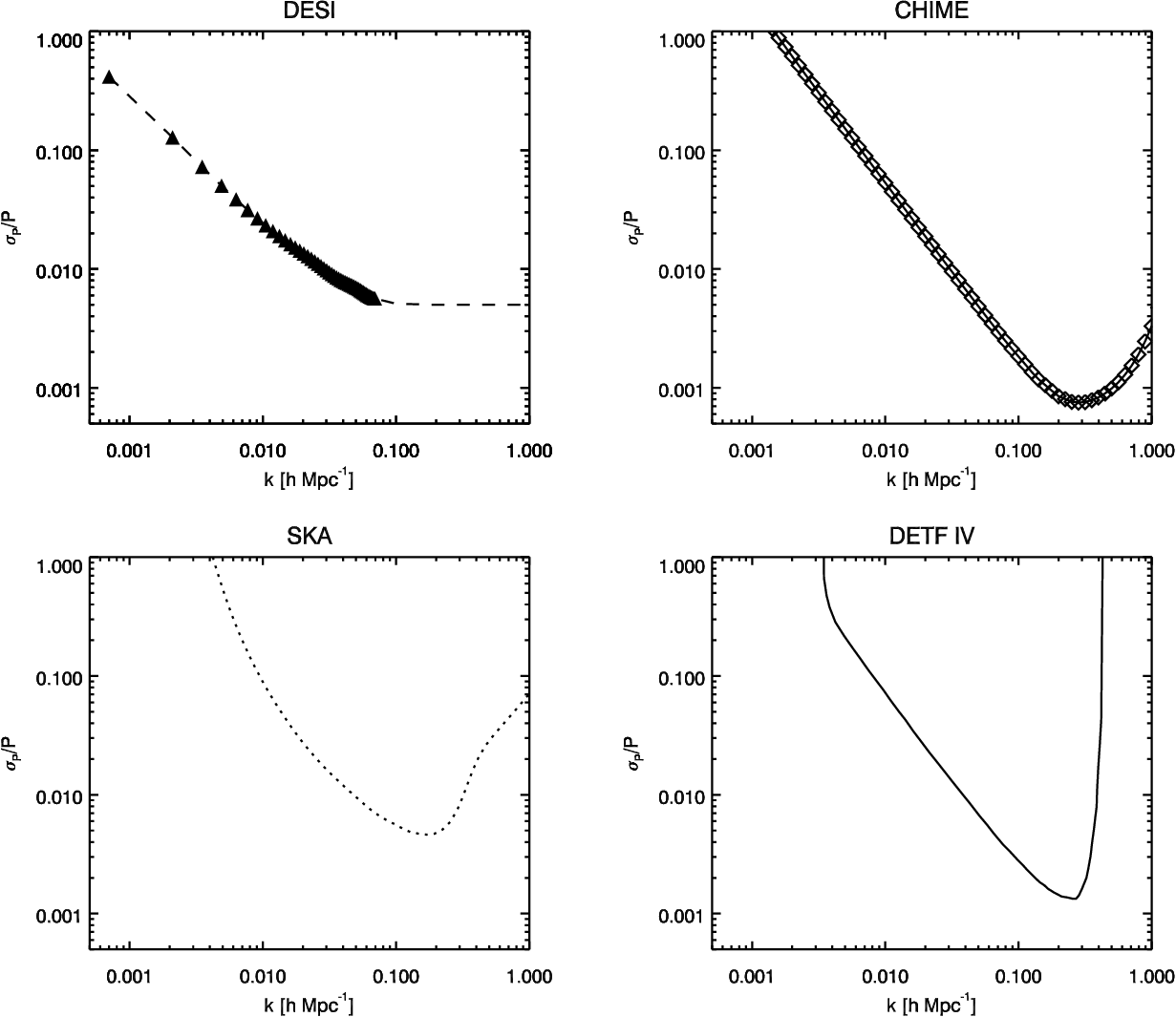}
\caption{ Comparison of different large-scale structure mapping surveys. The plot shows the fractional error in the power spectrum versus wave number $k$ for DESI, CHIME, SKA, and a stage IV Dark Energy Task Force experiment (DETF IV). We emphasise that this is not a direct comparison. Different experiments use different schemes for binning in $k$ and different redshift intervals and target redshifts. Here we merely show a snapshot of the capabilities of different surveys.
The Dark Energy Spectroscopic Instrument (DESI, \url{http://www.desi.lbl.gov}) is a galaxy and quasar/Lyman-alpha survey that will measure the effect of dark energy on the expansion of the universe.  It will obtain optical spectra for tens of millions of galaxies and quasars. The DESI estimated errors were provided by Pat McDonald (2014, priv. comm.). The DESI points are truncated to 0.5 per cent error for the small scales or larger  $k$ as a rough estimate of potential systematic effects,
which are essentially neglected in the other surveys at this point. 
The Canadian Hydrogen Intensity Mapping Experiment (CHIME) is a 21 cm survey \citep{CHIME:2013}. The CHIME errors, calculated using the method in \citet{Seo:2010}, were provided by Kris Sigurdson (private communication).
The DETF IV (survey aims specifically at measuring the BAO features very accurately to test for dark energy effects, but does not focus on testing the very-high-energy effects of inflation, namely, looking at the horizon scale at recombination and above, where the effects
of early inflation and early universe phase transitions are most likely to be observed. The SKA (phase 1) and DETF fractional errors are obtained from \citet{Bull:2014}.
}

\label{fig:allsurveys}
\end{figure*}

\subsection{Comparison with other surveys}
It is relevant to compare the possible FAST surveys with other proposed surveys. Such a comparison is not without difficulty. Forecasts for different surveys use different assumptions. The accuracy of the forecasts depends considerably on detailed knowledge of the final design specifications. Again, this varies for different surveys. However, for the scope of this paper it is still pertinent to compare the different instruments to obtain a first-order measurement of the capabilities of FAST in relation to other surveys.

\begin{figure*}
\centering
\includegraphics[angle=270,width=12cm]{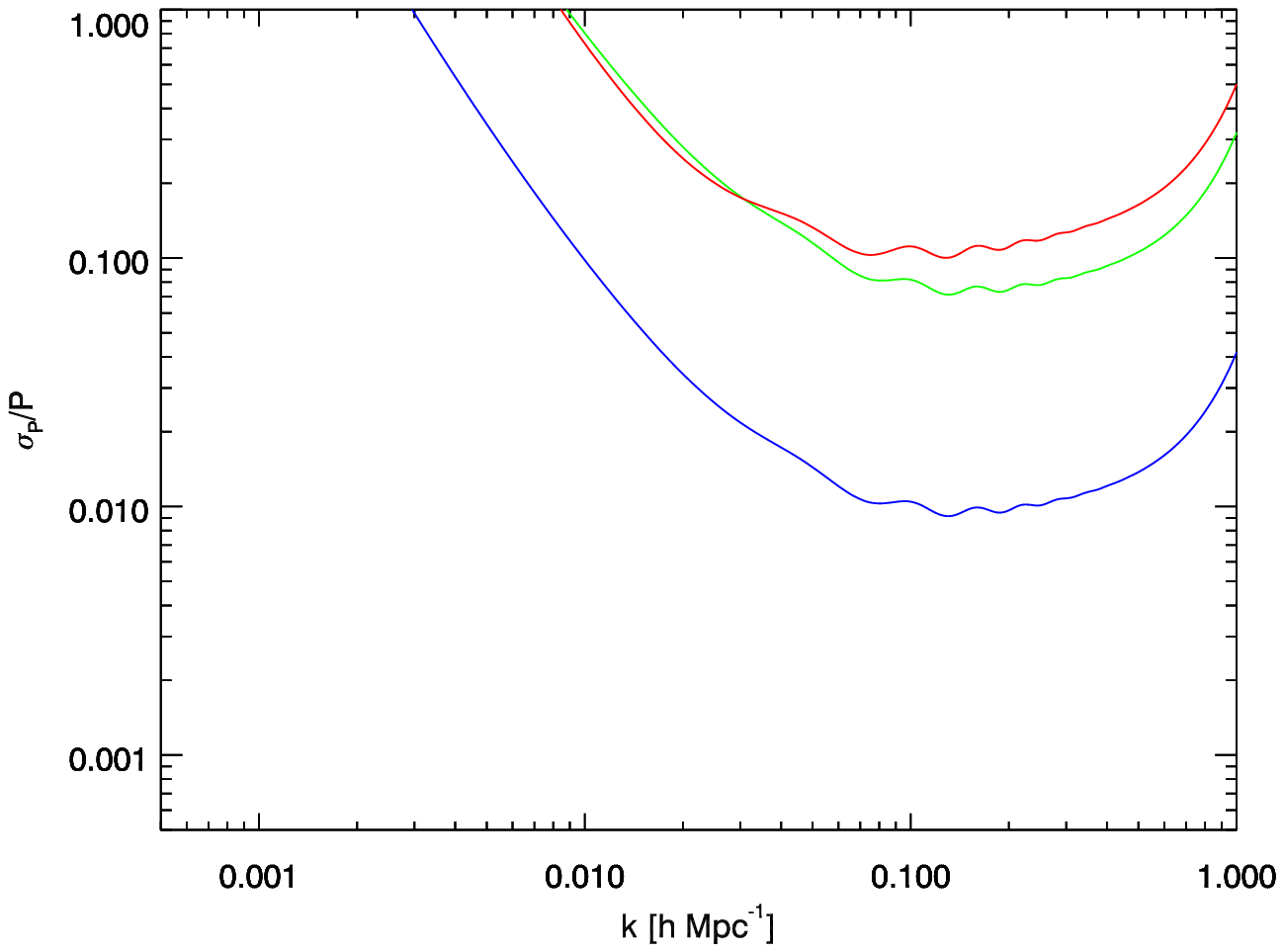}
\caption{Different $\Delta k$ binning for the FAST surveys. The plot shows the fractional error in the power spectrum versus wave number $k$ at a redshift of $z = 0.9$ for the FAST surveys. All the other survey parameters are the same as for Fig. \ref{fig:surveys}. Here we use a binning scheme for FAST where $\Delta k = k$ (see Eq. \ref{eqn:error}). The different FAST survey configurations are green for survey 1, red for survey 2, and blue for survey 2 with two years of observations.}

 \label{fig:allsurveysk3} 
\end{figure*}

\begin{figure*}
\centering
\includegraphics[angle=90,width=12cm]{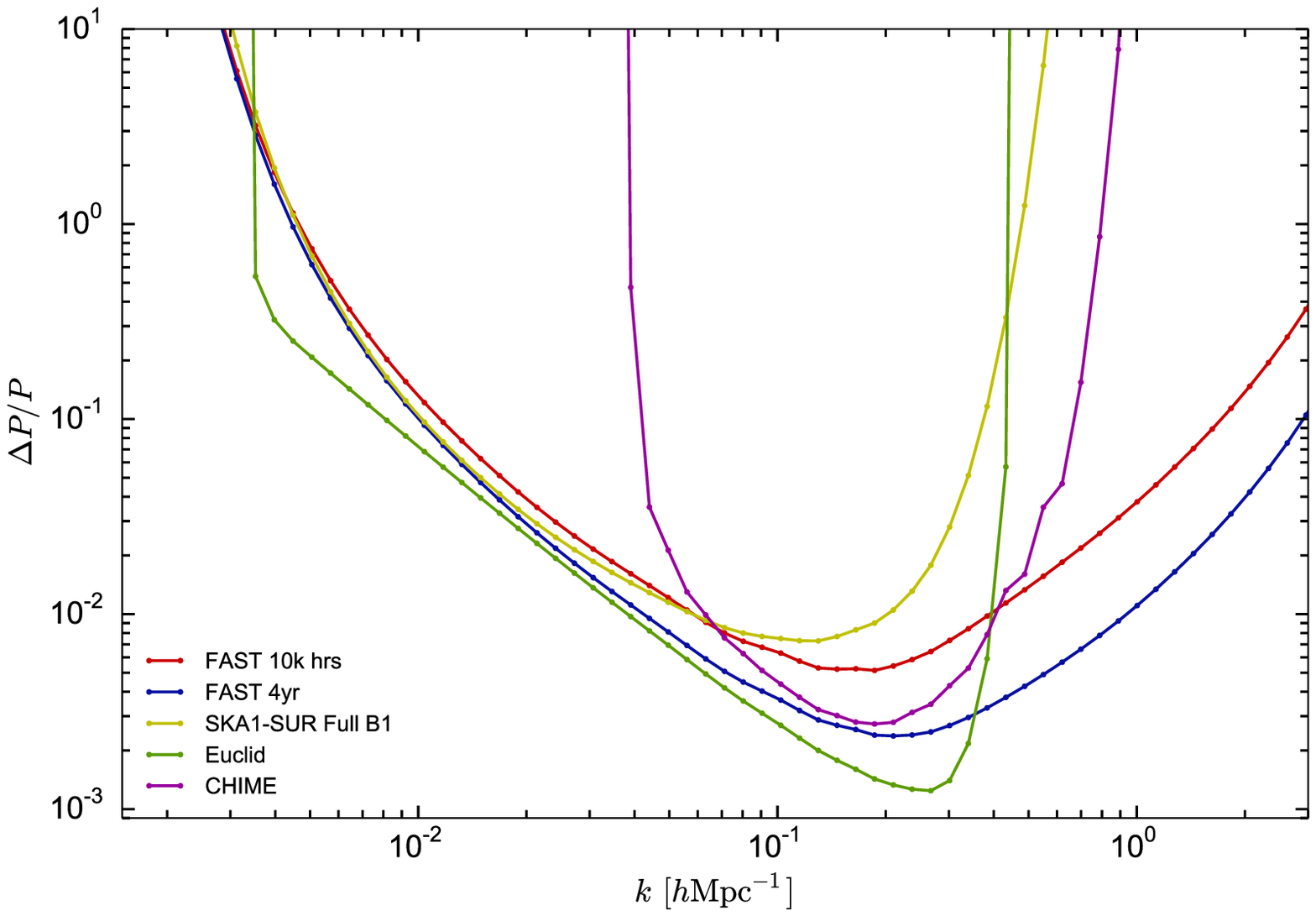}
\caption{  Yet another comparison of surveys, all with sky coverage of about 25 000 square degrees. 
The plot shows the fractional error in the power spectrum versus wave number $k$
and shows that the FAST and SKA1 survey (phase 1 of the SKA) are more directly comparable. SKA1-MID will carry out an intensity mapping survey, but SKA1-SUR is no longer planned \citep{Santos2015,Bull2016}.
The curve labelled Euclid  effectively represents cosmic variance-limited observations over a similar redshift range. The survey has the same specifications as the DETF IV survey shown in the previous plots. This shows that the two programmes can approach cosmic variance limits in the central $k$ range
but begin to suffer from limitations in the signal-to-noise ratio at low $k < 10^{-2}\,h {\mr{Mpc}}^{-1}$ and $k > 0.2\,h {\mr{Mpc}}^{-1}$
This plot was provided by Philip Bull (private communication). It follows the method used in \citet{Bull:2014}. All surveys use the same binning scheme: 20 bins per $k-$decade. 20 feeds are assumed for the two FAST surveys.  The redshift ranges are as follows: the two FAST surveys (0.42 to 2.55), SKA1-SUR (0.58 to 2.55), Euclid (0.65 to 2.05), and CHIME (0.77 to 2.55). The CHIME sensitivity shown in Figs. \ref{fig:allsurveys} and \ref{fig:allsurveysk3} is calculated using a different technique to the one used here.}
\label{fig:allsurveysBull}
\end{figure*}

Figure \ref{fig:allsurveys}  shows the nominal fractional sensitivity of various proposed surveys, quoted in different publications. This comparison is difficult because all the different proposed surveys quote their sensitivity using different binning schemes in $k$. 
We therefore show another plot in Fig. \ref{fig:allsurveysk3} where the FAST surveys have simple $\Delta k = k$ bins and 
the special Fig. \ref{fig:allsurveysBull}, which has all surveys with the same 20 bins per decade, 
kindly provided by Philip Bull (see \citealt{Bull:2014}).

\subsection{Motivated models of large-scale perturbations}
The first anticipated science output of FAST is in testing models of large-scale perturbations. 
In the concordance model of cosmology, the growth of structure is seeded by primordial quantum perturbations that depend on some inflationary potential. 
To date, the model has used the slow-roll potential model, where the primordial power spectrum model is parameterised by a power law of the form
 \bq P_\mr{S}^{\mr{Plaw}}(k)=A_s\left(\frac{k}{k_{0}}\right)^{n_\mr{s}-1},\eq  
where $A_s$ is the normalisation, $k_0$ is the pivot scale and $n_\mr{s}$ is the tilt. 
The latter is $n_s=1$ for a scale-independent spectrum.  

The most remarkable result to come out of the \textit{Planck} mission so far is the $5\sigma$ statistical difference between the data and the scale-independent primordial power spectrum model \citep{Planck-Collaboration:2013aa,Planck-Collaboration2015aa}.
This strengthens and verifies the observed `low power at large scales' anomaly first observed by 
COBE \citep{COBE:1992,Hinshaw:1996,Bond:1998} and then further comfirmed by the  WMAP results \citep{WMAP:2003}.

These observations lead us to conclude that the primordial power spectrum is scale dependent. 
The recent claim by the BICEP2 team \citep{BICEP} of the detection of primordial gravitational waves has reopened the crucial question of inflation, motivating the proposition of various models. 

The Wiggly Whipped Inflation model \citep{Hazra:2014} seems to fit joint \textit{Planck} and BICEP-2 data (particularly the deviation from a smooth primordial power spectrum) better than other models. 
These fall in the general class of `just enough inflation' models \citep{Cicoli:2014},
which generically produce suppression of power on the largest angular scales and an oscillatory  behaviour at scales around the Hubble parameter at the beginning of the slow roll inflation that covers most of the observations of CMB and large-scale structure.

The primordial power spectrum resulting from this model leaves an imprint on the large-scale structure at the present epoch. 
More importantly, it may show features at scales that can be probed by weak-lensing surveys such as the future 
Euclid mission.

One version of the Wiggly Whipped Inflation potential, referred to as second order in 
\citet{Hazra:2014}, has the form
\bq V(\phi)=\gamma \phi^2+\lambda\phi^2(\phi-\phi_0)\Theta(\phi-\phi_0) ,\eq 
where $\Theta$ is a Heavyside step function  modified to be the smooth function
\bq\Theta(\phi-\phi_0)= \frac{1}{2}\left[1 + \tanh\left(\frac{\phi - \phi_0}{\Delta\phi}\right)\right]~,\eq
where $\Delta \phi$ is a constant indicating the smoothness of the step and 
$\phi_0$ is the transition point of the inflaton field.  

To measure the features in the matter power spectrum that would be imprinted by this inflation potential, we need to determine any deviations from a power-law fit to the data. We therefore considered the ratio of the power spectra given by Wiggly Whipped Inflation and the simplest power-law inflation.

We calculated the full non-linear matter power spectrum $P(k)$ using the publicly available Boltzmann code \textsc{camb}
  \citep{CAMB}, taking the \textit{Planck} best-fit values for our fiducial cosmology: 
 $h=0.7,\quad \Omega_\mr{\Lambda}=0.7,\quad  w=-1,\quad \Omega_\mr{m} = 0.3,\quad \Omega_\mr{b}= 0.0462,\quad 
 \tau= 0.09,\quad N_\mr{eff}= 3.046$. The power-law inflation parameterisation was set to \textit{Planck} best-fit values, with {$n_s=0.96$} and $k_0=0.05\,\mr{Mpc}^{-1}$. 

The Wiggly Whipped Inflation potential was calculated using the numerical code \textsc{bingo}  \citep{Hazra:2013}. 
We chose fiducial values that provided the best fit to a combination
of \textit{Planck} + WP + BICEP2 data. 
Thus we set $\gamma=2.68\times10^{-11}$, $\lambda=5.2\times10^{-13}$ and $\phi_0=14.59 \,\mr{M_{\mr{Planck}}}$.

\begin{figure*}
\centering
{\includegraphics[angle=270,width=12cm]{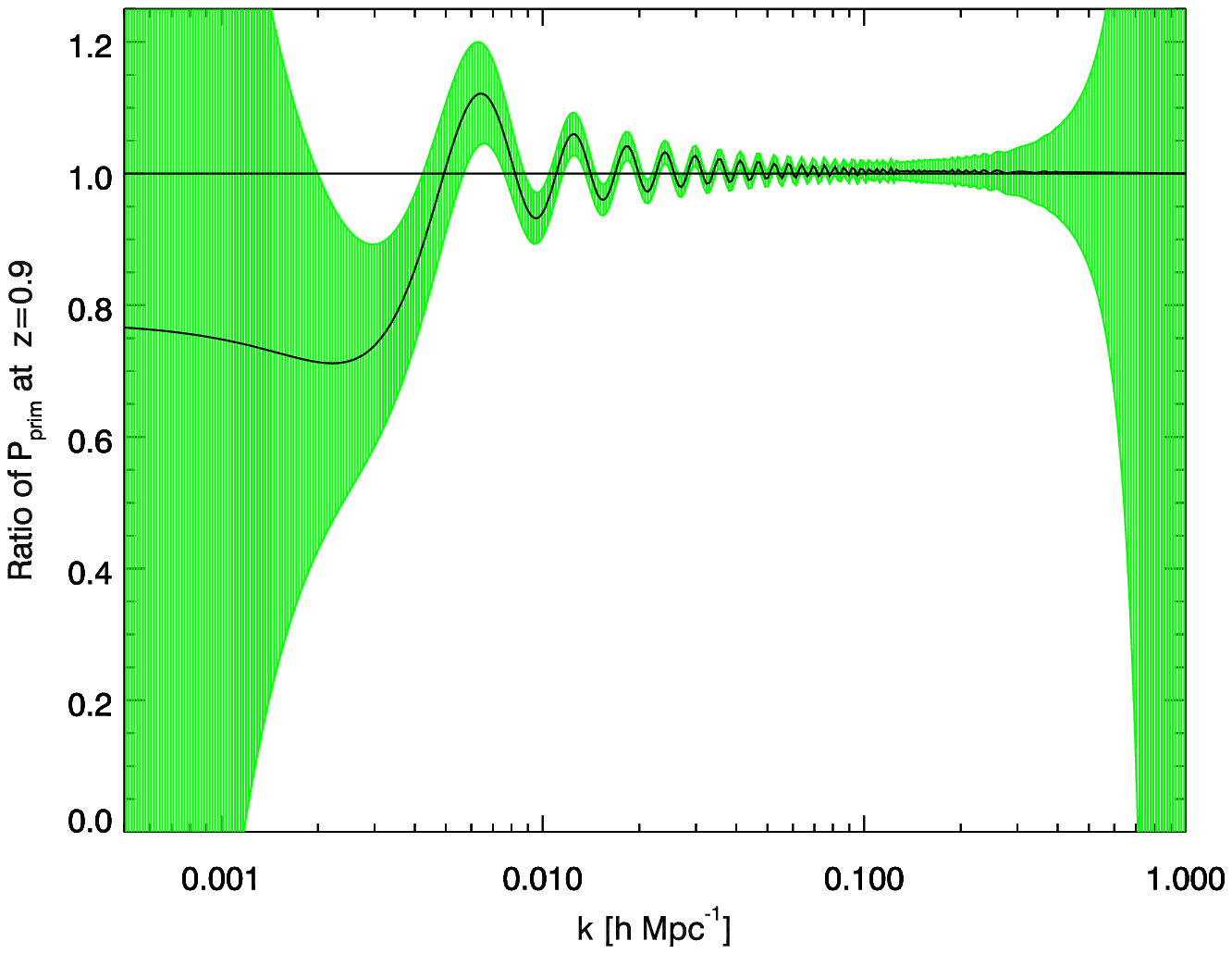}}
\caption{Large-scale perturbation power spectrum with error bands for survey 1 . The y-axis is the ratio of the power-law primordial to a Wiggly Whipped primordial scalar power spectrum. The errors are calculated for a survey redshift $z=0.9$.  }
\label{fig:Wiggly1}
\end{figure*}

\begin{figure*}
\centering
\includegraphics[angle=270,width=12cm]{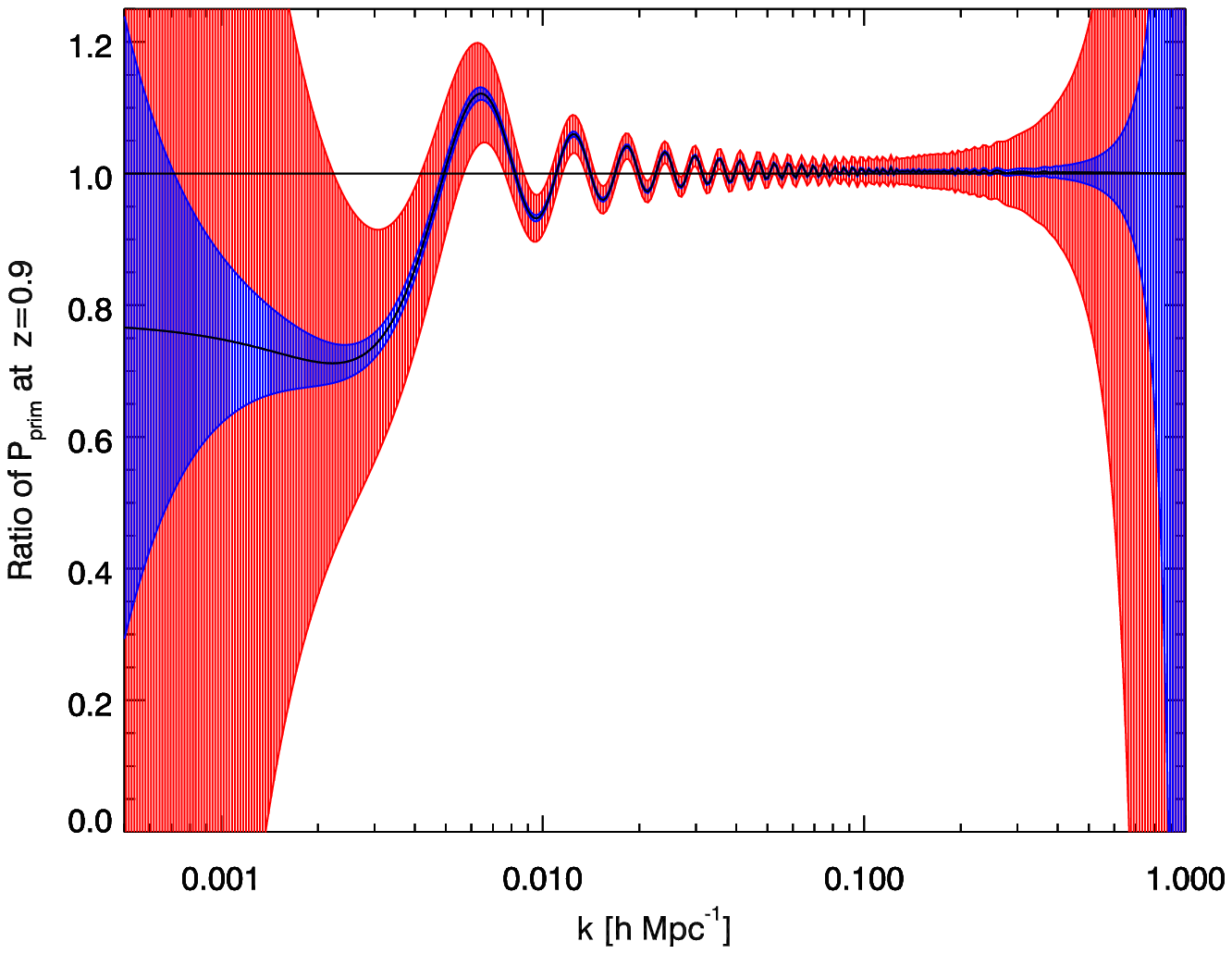}
\caption{ Large-scale perturbation power spectrum with error bars for survey 2, the 3D 21 cm intensity map. There are two sets of error contours. The large ones (red) are for a 48-day survey. The inner smaller errors (blue) are for an effective two-year survey.}
\label{fig:Wiggly2} 
\end{figure*}

In Figs. \ref{fig:Wiggly1} and \ref{fig:Wiggly2} we show the errors on this ratio given by the FAST survey considered in this paper for second-order Wiggly Whipped Inflation \citep{Hazra:2014}. 
It is important to emphasise here that models with just enough inflation (50 to 60 e-folds) will generally show
features in the regime $0.0001 < k < 0.002$ and then wiggles beginning around $k \sim 0.002$ Mpc$^{-1}$ and damping to larger $k$.
Observations bias the low $k$ features on average to be a dip.

\begin{figure*}
\centering
\includegraphics[angle=270,width=12cm]{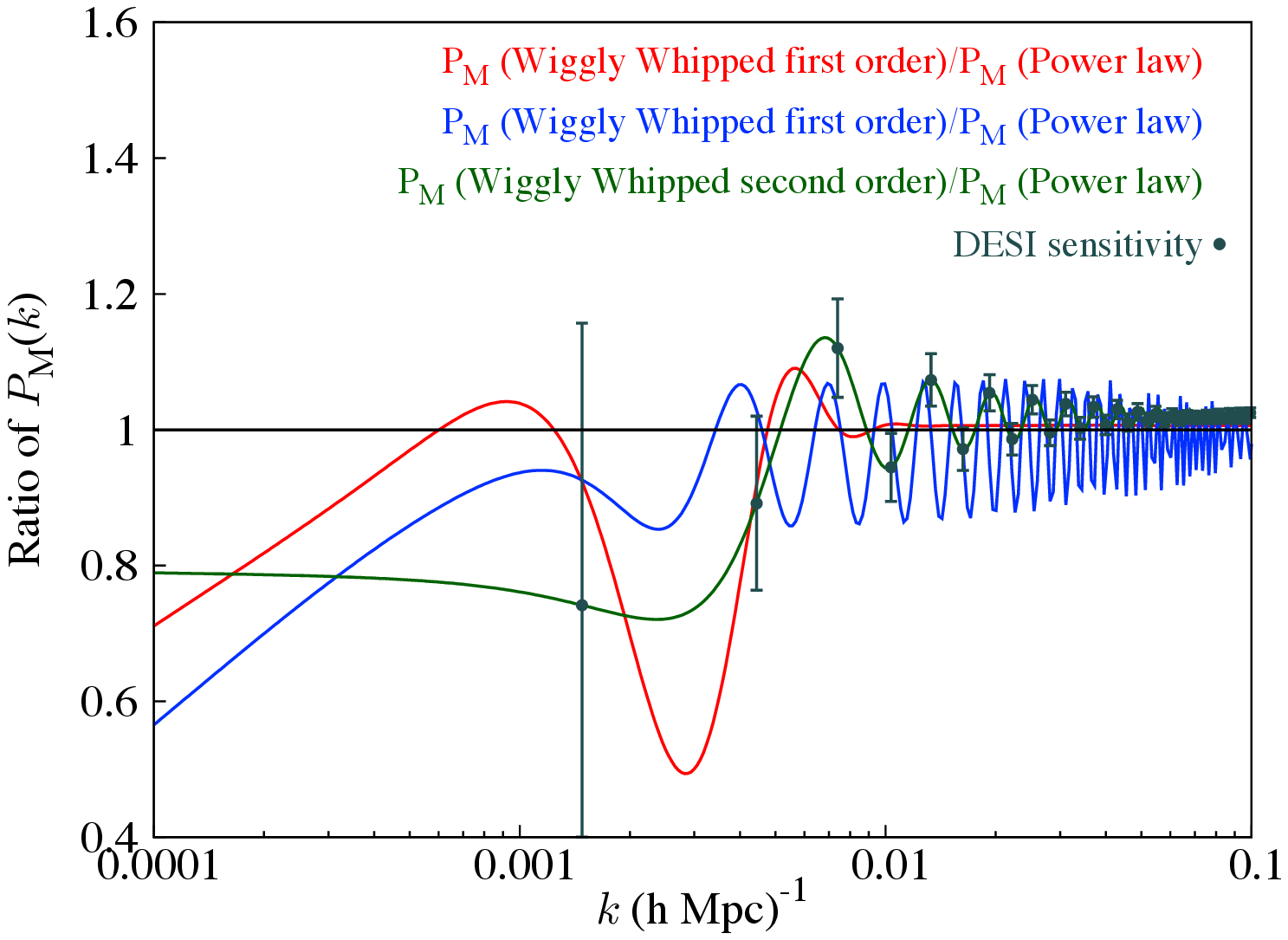}
\vskip -0.5in
\includegraphics[angle=270,width=12cm]{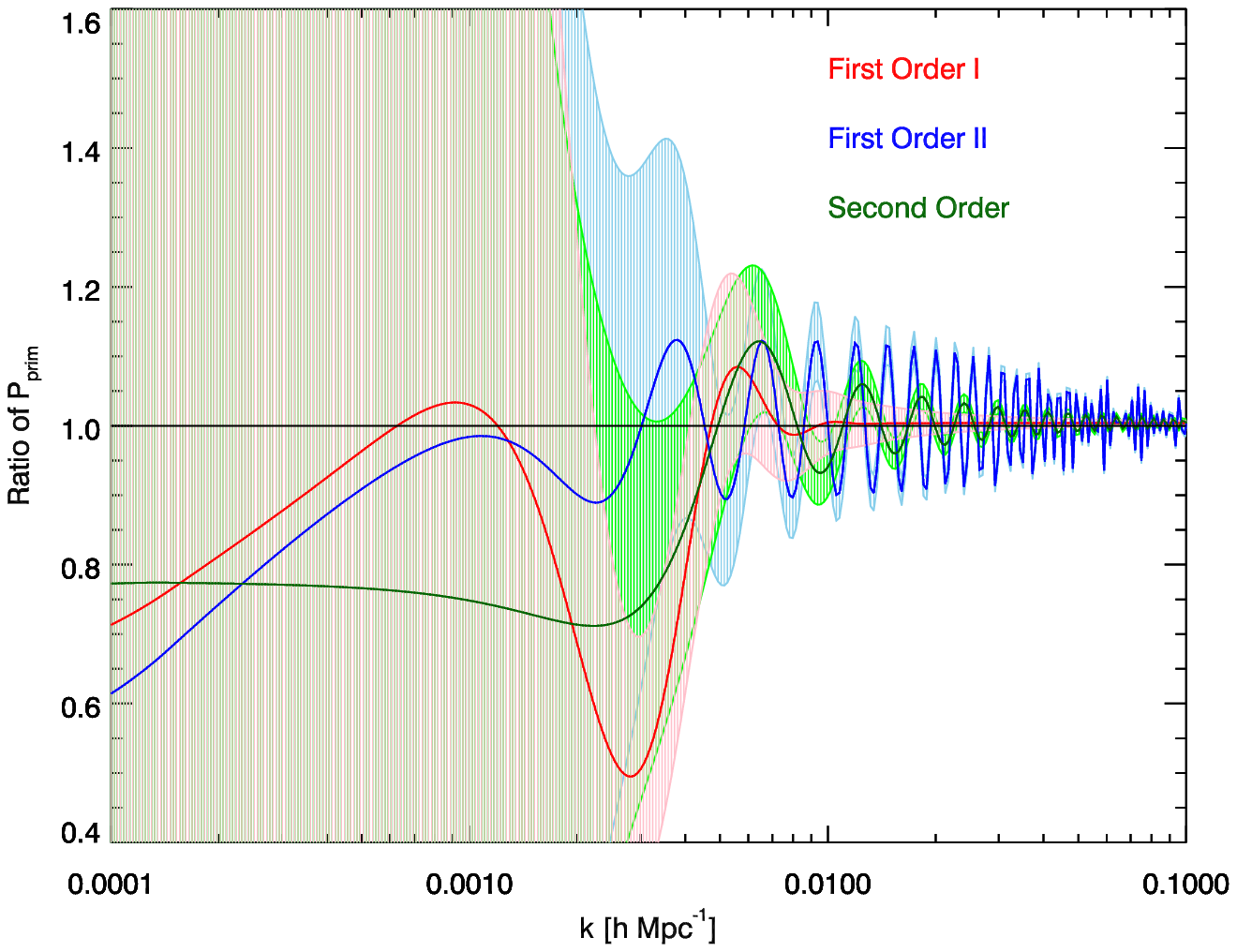}
\caption{Example of the type of large-scale primordial perturbations 
expected for large field inflation affected by the GUT transition or other features. 
These example have the generic features of `just enough' Inflation (50-60 $e$-folds). 
The three curves are three versions of Wiggly Whipped Inflation \citep{Hazra:2014}, best fit to CMB data: first-order version I (red), first-order version II (blue), and second order (green). The latter is the particular model used in the previous plots.
The top panel shows the error bars for DESI (private communication from Pat McDonald) on the 
second-order Wiggly Whipped Inflation model. The lower panel shows the same inflation models, 
with the associated error bands (for all three curves) for an idealised FAST 2 survey extended to four years (survey 2c), with the sky coverage doubled.  
Note that 21 cm intensity mapping on large scales is an ideal efficient method to obtain high-quality data in this region.}
 \label{fig:Inflation} 
\end{figure*}

Figure \ref{fig:Wiggly1} shows the perturbation power spectrum from Wiggly Whipped Inflation with the error bands for survey 1 (basically a 2D 21 cm intensity map). Figure \ref{fig:Wiggly2} shows the same ratio for survey 2.  In Fig. \ref{fig:Inflation} we show this ratio with three versions of the Wiggly Whipped Inflation potential for an idealised survey 2 (called survey 2c in  Table \ref{tab:surveys}).

Figures \ref{fig:Wiggly2} and \ref{fig:BAO2} show the type of performance expected for the initial 3D scan and an extensive two-year observing programme.

The forecasts presented here illustrate the capabilities of FAST in studying the large-scale perturbation power spectrum, which is only one of the many science objectives of the survey. We note that the quality of any future results depends on the modelling of aspects such as the foregrounds, 1/f noise, the effect of sidelobes, and contamination. This effort requires time and resources.

\subsection{Observing BAO features}
Another clear basic science output for FAST 21 cm intensity mapping is the observation of BAO features and the determination of cosmological parameters. The results can also be compared to other survey techniques, including galaxy and weak-lensing surveys. 

\begin{figure*}
\centering 
\includegraphics[angle=270,width=12cm]{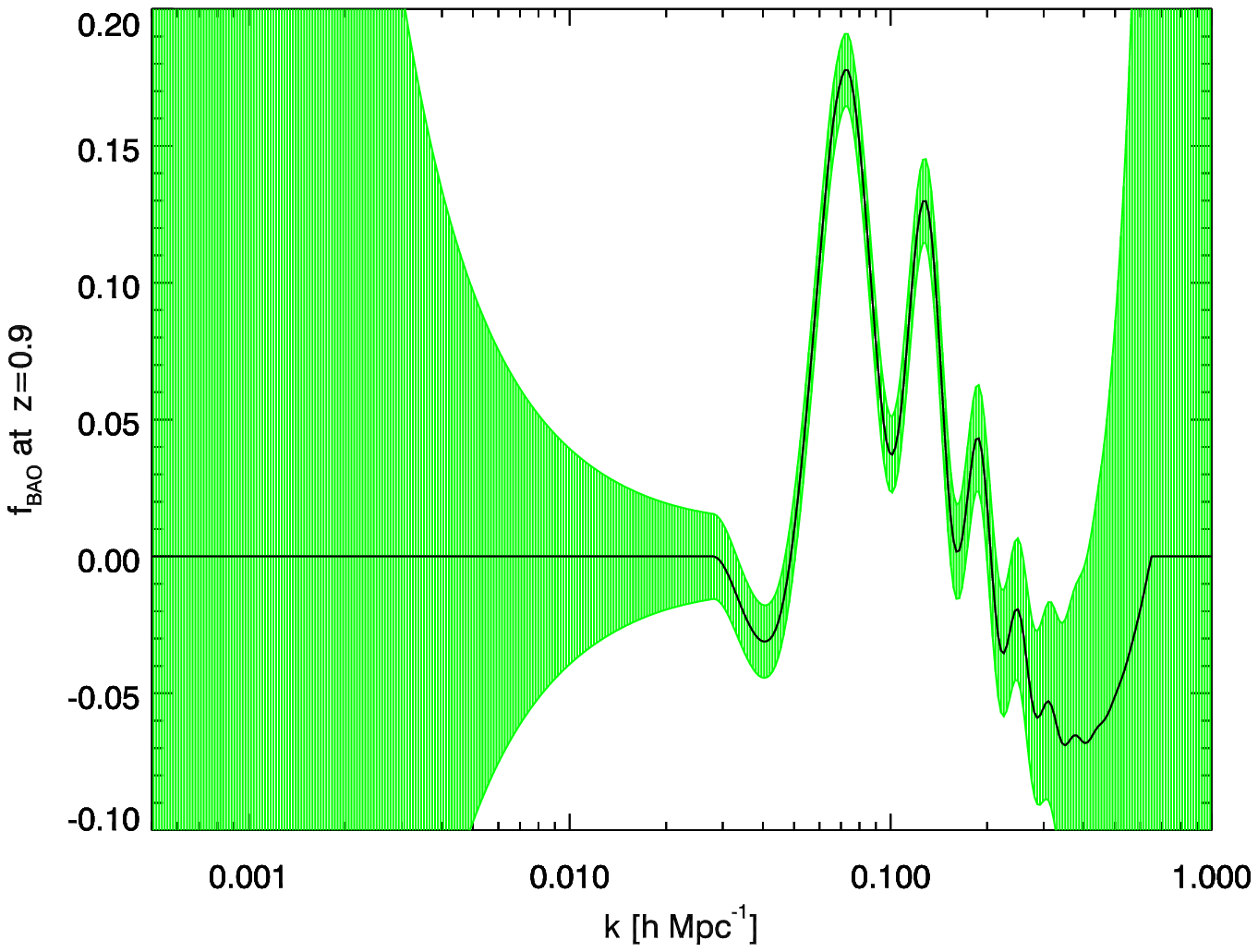}
\caption{ Focus on BAO features: perturbation power spectrum with error bands for survey 1, which is basically a 2D 21 cm intensity map. $f_{\mr{BAO}}$ is the ratio of the processed with BAO power spectrum to a power-law spectrum plus simple power-law cut-off for the processing suppression. }
\label{fig:BAO1}
\end{figure*}

\begin{figure*}
\centering
\includegraphics[angle=270,width=12cm]{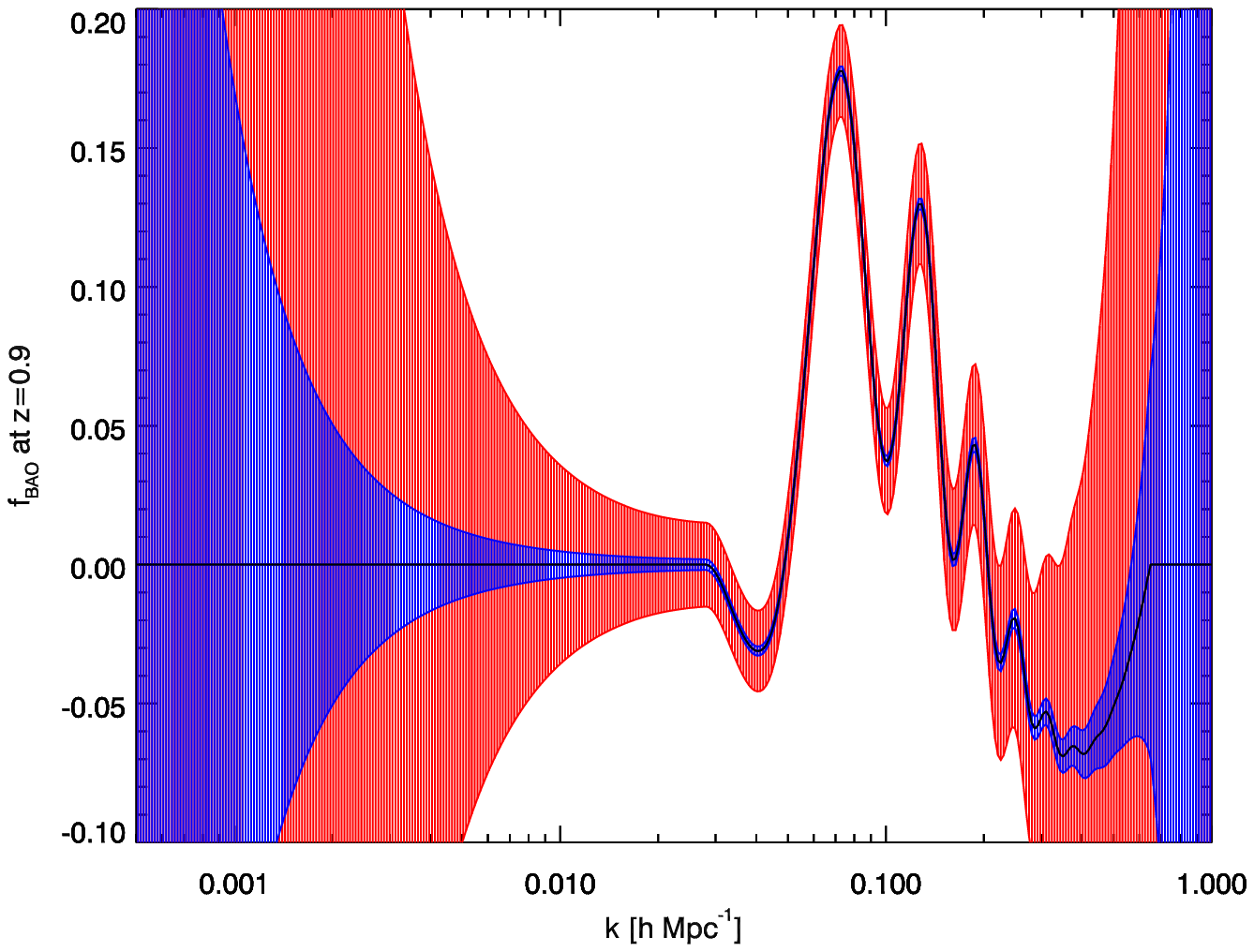}
\caption{Focus on BAO features: the perturbation power spectrum with error bands for 
survey 2, which is the 3D 21 cm intensity map. The outer (red) errors are for the nominal 48-day observing first pass (survey 2a). 
The inner (blue) error band is for a two-year observing run (survey 2b) assuming that the data integrate down and full
 sensitivity is achieved and thus the errors on the power spectrum decrease inversely with observing time.}
 \label{fig:BAO2} 
\end{figure*}

In Figs. \ref{fig:BAO1} and \ref{fig:BAO2} we show the error bands for surveys 1 and 2 for $f_{\mr{BAO}}$. This is the ratio of the power spectrum processed with BAO to a power-law spectrum plus a simple power-law cut-off for the processing suppression. $f_{\mr{BAO}}$ is forced to zero outside the $k$ range of interest 
so as not to consider the possible low-$k$ perturbations and the more non-linear effects at large $k$. The error bands are given by $f_{\mr{BAO}} \pm \sigma_P / P$. The fractional errors $ \sigma_P / P$ are the anticipated statistical errors on the matter power spectrum.

\begin{figure*}
\centering
\includegraphics[angle=270,width=12cm]{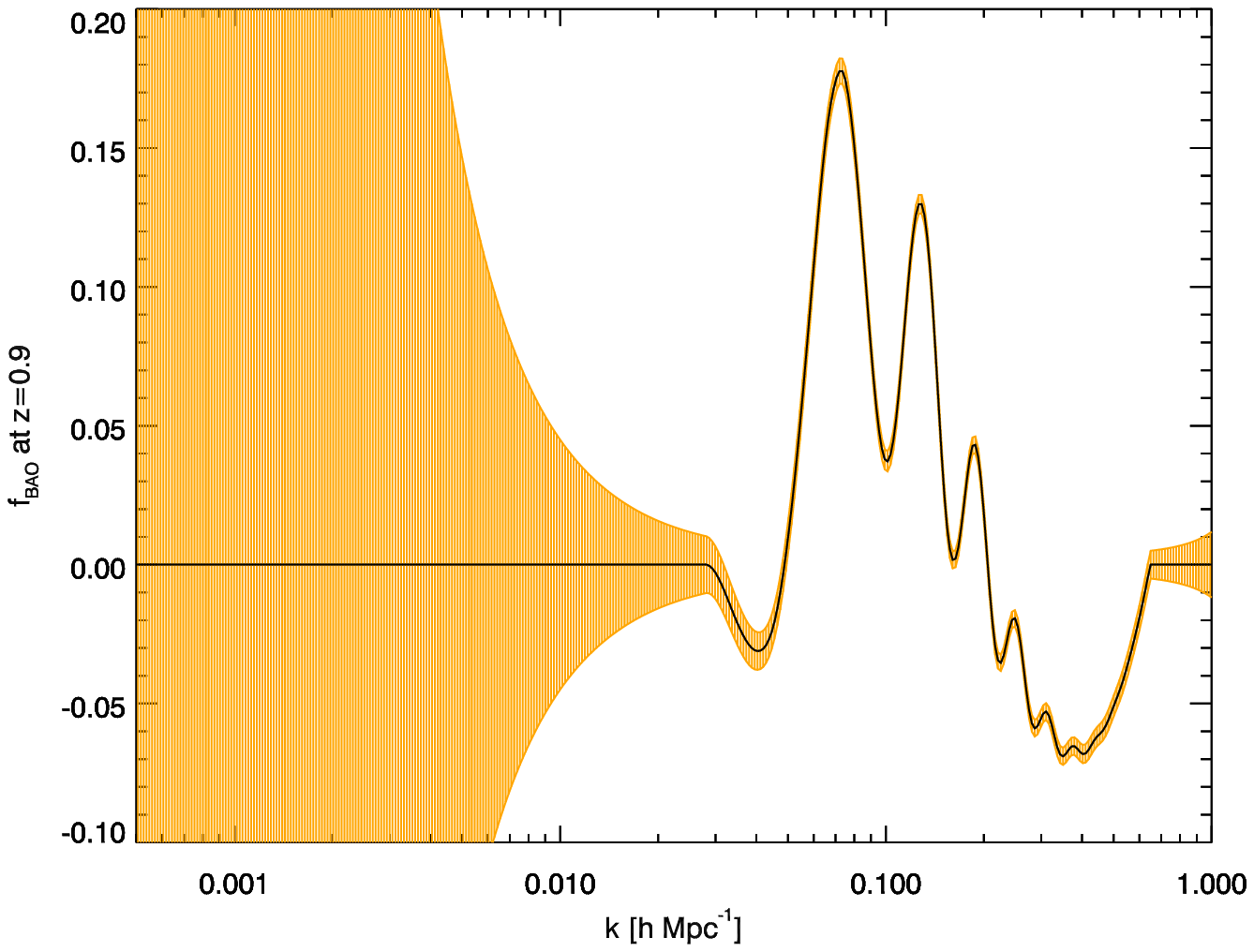}
\caption{Example of the type of BAO perturbations features with  $f_{\mr{BAO}} \pm \sigma_P / P$ error contours
 expected from FAST 21 cm intensity mapping. 
Here we use the idealised survey (survey 2c). }
\label{fig:BAO} 
\end{figure*}

Figure \ref{fig:BAO} shows the results that can be achieved with an idealised FAST survey  (survey 2c).

Measuring the BAO features with accuracy and precision and more than one technique allows us to probe 
the nature of dark energy (the acceleration of the Universe), large-scale structure formation, dark matter 
effects and properties, and to test alternative theories of gravity. 
As shown here and in the two previous BAO figures, the FAST 21
cm intensity mapping has the potential of measuring these features well in the critical and scientifically powerful redshift range from $0.5 < z < 2.5$. 
These maps can be
extended and improved should the scientific results warrant it.

\subsection{Anticipated ancillary science}

Because of the unprecedented combination of a large collecting area, good angular resolution, and 24-hour operation, 
FAST will detect and monitor thousands of previously unknown transient sources. 
The bulk of radio transients that have been discovered to date have been found using the Parkes Multi-beam survey\footnote{\url{http://www.atnf.csiro.au/research/multibeam}} (11 so far). Because of the small number of sources thus far observed, the luminosity function is very poorly constrained, 
and therefore precise projections of the detection rate are not possible. 

FAST observations will reach much deeper than the Parkes survey.
If both CHIME and the FAST 21 cm survey are operating at the same time,
then sharing catalogues and observations of these sources will make both surveys more productive in this effort.
CHIME has the advantage that it is scanning a large celestial latitude band, simultaneously providing more coverage and detection rate,
while FAST has the advantage of a larger collecting area and thus signal-to-noise ratio on these events.
The FAST intensity mapping is a transit mode: once a source is detected, the pulse pattern of the source can be searched for, 
both in future scans and retrospectively, in each day's data. 
Given the recent success of radio-transient searches, 
it is reasonable to anticipate that completely new types of sources will appear in the
FAST observations.

Because of its large field of view, 
FAST will provide the capability to time known pulsars, and eventually to search for new ones, especially in combination with other surveys (see e.g. \citealt{FASTHobbs2014}).
FAST also has receivers in many bands, and a programme is in place for pulsar timing and dispersion/de-dispersion treatment.
All pulsars in the mostly northern hemisphere will spend from five minutes to hours within the FAST intensity mapping fields of view. 
From a list of locations and dispersion measures, a data set of timing and scintillation for hundreds of pulsars will be produced.

Finally, FAST has the potential of contributing to the knowledge and understanding of the Galactic magnetic field. The effort in this field is led by the GMIMS experiment. The main objective of GMIMS [Global Magneto-Ionic Medium Survey \citep{Wolleben:2009}] is to make a map of the diffuse polarised emission over the entire sky. 
GMIMS will do this by mapping the polarised radio emission from the Milky Way in the frequency band 300 to 1800~MHz. 

Data in this spectral range will reveal aspects of how magnetic  fields regulate star formation and couple the energy released by
stellar winds and supernovae to the interstellar medium. 
FAST, if polarisation information is included, will provide much more sensitive data over the middle part of this frequency band.
By including more channels or dual polarisation, we could measure the linear polarisation of the sky observations. Dual-polarisation feeds could provide both the intensity and polarisation components of the sky in switching or correlation mode. However, this would increase the cost of the project.

\section{Conclusions and future}

We have shown that FAST can be a useful addition to radio astronomy until the Square Kilometre Array becomes operational. It enables us to obtain good science results within a short time-frame. These results can then be tested and refined upon by the SKA. FAST can give good constraints on different inflation scenarios and on BAO physics. Although the Square Kilometre Array is expected to do better when the full experiment is operational, this is planned for the late 2020s, more than a decade after FAST starts providing science results.

The performance of FAST is comparable to CHIME. The two experiments have similar collecting areas. There are some differences in the number of photons observed, and in the ability to exclude the unwanted signal, but overall the potential capability is similar. 

The potential programme outlined here offers very strong science products.
They indicate that FAST has the potential to provide a treasure trove of cosmological observations.
There are a number of ways to improve our understanding of the universe and the potential of this instrument and this potential programme.
The next steps are to refine the survey parameters and observing plan
and to provide a full concept and design for the receiver and data processing systems.

\begin{acknowledgements}
G.F.S. acknowledges support through his Chaire d'Excellence Universit\'e 
Sorbonne Paris Cit\'e and the financial support of the UnivEarthS Labex programme at 
Universit\'e Sorbonne Paris Cit\'e (ANR-10-LABX-0023 and ANR-11-IDEX-0005-02). 
I.D. acknowledges that the research work disclosed in this publication is partially funded by the 
REACH HIGH Scholars Programme --- Post-Doctoral Grants. The grant is part-financed by the European Union, Operational Programme II ---
Cohesion Policy 2014--2020. The authors would also like to thank Pat McDonald, Richard Shaw, Kris Sigurdson, Matt Dobbs, Kevin Bandura, Phil Bull, Pedro Ferreira, Mario Santos, David Alonso and Dhiraj Kumar Hazra for useful discussions and feedback.
\end{acknowledgements}

\bibliographystyle{aa}


\begin{thebibliography}{55}
\expandafter\ifx\csname natexlab\endcsname\relax\def\natexlab#1{#1}\fi

\bibitem[{{Ade} {et~al.}(2014){Ade}, {Aikin}, {Barkats}, {Benton}, {Bischoff},
  {Bock}, {Brevik}, {Buder}, {Bullock}, {Dowell}, {Duband}, {Filippini},
  {Fliescher}, {Golwala}, {Halpern}, {Hasselfield}, {Hildebrandt}, {Hilton},
  {Hristov}, {Irwin}, {Karkare}, {Kaufman}, {Keating}, {Kernasovskiy}, {Kovac},
  {Kuo}, {Leitch}, {Lueker}, {Mason}, {Netterfield}, {Nguyen}, {O'Brient},
  {Ogburn}, {Orlando}, {Pryke}, {Reintsema}, {Richter}, {Schwarz}, {Sheehy},
  {Staniszewski}, {Sudiwala}, {Teply}, {Tolan}, {Turner}, {Vieregg}, {Wong},
  {Yoon}, \& {Bicep2 Collaboration}}]{BICEP}
{Ade}, P.~A.~R., {Aikin}, R.~W., {Barkats}, D., {et~al.} 2014, Phys. Rev.
  Lett., 112, 241101

\bibitem[{Alonso {et~al.}(2014)Alonso, Ferreira, \& Santos}]{Alonso:2014}
Alonso, D., Ferreira, P.~G., \& Santos, M.~G. 2014, MNRAS, 444, 3183

\bibitem[{{Banday} \& {Wolfendale}(1990)}]{Banday:1990}
{Banday}, A.~J. \& {Wolfendale}, A.~W. 1990, \mnras, 245, 182

\bibitem[{{Banday} \& {Wolfendale}(1991)}]{Banday:1991}
{Banday}, A.~J. \& {Wolfendale}, A.~W. 1991, \mnras, 248, 705

\bibitem[{{Bandura} {et~al.}(2014){Bandura}, {Addison}, {Amiri}, {Bond},
  {Campbell-Wilson}, {Connor}, {Cliche}, {Davis}, {Deng}, {Denman}, {Dobbs},
  {Fandino}, {Gibbs}, {Gilbert}, {Halpern}, {Hanna}, {Hincks}, {Hinshaw},
  {H{\"o}fer}, {Klages}, {Landecker}, {Masui}, {Mena Parra}, {Newburgh}, {Pen},
  {Peterson}, {Recnik}, {Shaw}, {Sigurdson}, {Sitwell}, {Smecher}, {Smegal},
  {Vanderlinde}, \& {Wiebe}}]{CHIME:2013}
{Bandura}, K., {Addison}, G.~E., {Amiri}, M., {et~al.} 2014, in \procspie, Vol.
  9145, Ground-based and Airborne Telescopes V, 914522

\bibitem[{{Battye} {et~al.}(2012){Battye}, {Brown}, {Browne}, {Davis},
  {Dewdney}, {Dickinson}, {Heron}, {Maffei}, {Pourtsidou}, \&
  {Wilkinson}}]{Battye:2012}
{Battye}, R.~A., {Brown}, M.~L., {Browne}, I.~W.~A., {et~al.} 2012, e-print
  (ArXiv:1209.1041)

\bibitem[{{Battye} {et~al.}(2013){Battye}, {Browne}, {Dickinson}, {Heron},
  {Maffei}, \& {Pourtsidou}}]{Battye:2013}
{Battye}, R.~A., {Browne}, I.~W.~A., {Dickinson}, C., {et~al.} 2013, \mnras,
  434, 1239

\bibitem[{{Bennett} {et~al.}(2003){Bennett}, {Halpern}, {Hinshaw}, {Jarosik},
  {Kogut}, {Limon}, {Meyer}, {Page}, {Spergel}, {Tucker}, {Wollack}, {Wright},
  {Barnes}, {Greason}, {Hill}, {Komatsu}, {Nolta}, {Odegard}, {Peiris},
  {Verde}, \& {Weiland}}]{Bennett:2003}
{Bennett}, C.~L., {Halpern}, M., {Hinshaw}, G., {et~al.} 2003, \apjs, 148, 1

\bibitem[{{Bond} {et~al.}(1998){Bond}, {Jaffe}, \& {Knox}}]{Bond:1998}
{Bond}, J.~R., {Jaffe}, A.~H., \& {Knox}, L. 1998, \prd, 57, 2117

\bibitem[{{Bridle}(1967)}]{Bridle:1967}
{Bridle}, A.~H. 1967, \mnras, 136, 219

\bibitem[{{Bull}(2016)}]{Bull2016}
{Bull}, P. 2016, \apj, 817, 26

\bibitem[{{Bull} {et~al.}(2015){Bull}, {Ferreira}, {Patel}, \&
  {Santos}}]{Bull:2014}
{Bull}, P., {Ferreira}, P.~G., {Patel}, P., \& {Santos}, M.~G. 2015, \apj, 803,
  21

\bibitem[{{Camera} {et~al.}(2013){Camera}, {Santos}, {Ferreira}, \&
  {Ferramacho}}]{Camera:2013}
{Camera}, S., {Santos}, M.~G., {Ferreira}, P.~G., \& {Ferramacho}, L. 2013,
  Physical Review Letters, 111, 171302

\bibitem[{{Cane}(1979)}]{Cane:1979}
{Cane}, H.~V. 1979, \mnras, 189, 465

\bibitem[{{Chang} {et~al.}(2010){Chang}, {Pen}, {Bandura}, \&
  {Peterson}}]{HI:2000}
{Chang}, T.-C., {Pen}, U.-L., {Bandura}, K., \& {Peterson}, J.~B. 2010, \nat,
  466, 463

\bibitem[{{Cicoli} {et~al.}(2014){Cicoli}, {Downes}, {Dutta}, {Pedro}, \&
  {Westphal}}]{Cicoli:2014}
{Cicoli}, M., {Downes}, S., {Dutta}, B., {Pedro}, F.~G., \& {Westphal}, A.
  2014, \jcap, 12, 030

\bibitem[{{Dong} \& {Han}(2013{\natexlab{a}})}]{Dong2013a}
{Dong}, B. \& {Han}, J.~L. 2013{\natexlab{a}}, Pub. Astron. Soc. Australia, 30

\bibitem[{{Dong} \& {Han}(2013{\natexlab{b}})}]{Dong2013b}
{Dong}, B. \& {Han}, J.~L. 2013{\natexlab{b}}, Pub. Astron. Soc. Australia, 30

\bibitem[{{Duffy} {et~al.}(2008){Duffy}, {Battye}, {Davies}, {Moss}, \&
  {Wilkinson}}]{Duffy:2008}
{Duffy}, A.~R., {Battye}, R.~A., {Davies}, R.~D., {Moss}, A., \& {Wilkinson},
  P.~N. 2008, \mnras, 383, 150

\bibitem[{{Fixsen} {et~al.}(2011){Fixsen}, {Kogut}, {Levin}, {Limon}, {Lubin},
  {Mirel}, {Seiffert}, {Singal}, {Wollack}, {Villela}, \&
  {Wuensche}}]{ARCADE2skybrightness2}
{Fixsen}, D.~J., {Kogut}, A., {Levin}, S., {et~al.} 2011, \apj, 734, 5

\bibitem[{{Haslam} {et~al.}(1981){Haslam}, {Klein}, {Salter}, {Stoffel},
  {Wilson}, {Cleary}, {Cooke}, \& {Thomasson}}]{Haslam:1981}
{Haslam}, C.~G.~T., {Klein}, U., {Salter}, C.~J., {et~al.} 1981, \aap, 100, 209

\bibitem[{{Haslam} {et~al.}(1982){Haslam}, {Salter}, {Stoffel}, \&
  {Wilson}}]{Haslam:1982}
{Haslam}, C.~G.~T., {Salter}, C.~J., {Stoffel}, H., \& {Wilson}, W.~E. 1982,
  \aaps, 47, 1

\bibitem[{{Haverkorn} {et~al.}(2003){Haverkorn}, {Katgert}, \& {de
  Bruyn}}]{Haverkorn:2003}
{Haverkorn}, M., {Katgert}, P., \& {de Bruyn}, A.~G. 2003, \aap, 403, 1031

\bibitem[{{Hazra} {et~al.}(2014){Hazra}, {Shafieloo}, {Smoot}, \&
  {Starobinsky}}]{Hazra:2014}
{Hazra}, D.~K., {Shafieloo}, A., {Smoot}, G.~F., \& {Starobinsky}, A.~A. 2014,
  \jcap, 8, 48

\bibitem[{{Hazra} {et~al.}(2013){Hazra}, {Sriramkumar}, \&
  {Martin}}]{Hazra:2013}
{Hazra}, D.~K., {Sriramkumar}, L., \& {Martin}, J. 2013, \jcap, 5, 26

\bibitem[{{Hinshaw} {et~al.}(1996){Hinshaw}, {Branday}, {Bennett}, {Gorski},
  {Kogut}, {Lineweaver}, {Smoot}, \& {Wright}}]{Hinshaw:1996}
{Hinshaw}, G., {Branday}, A.~J., {Bennett}, C.~L., {et~al.} 1996, \apjl, 464,
  L25

\bibitem[{{Hobbs} {et~al.}(2014){Hobbs}, {Dai}, {Manchester}, {Shannon},
  {Kerr}, {Lee}, \& {Xu}}]{FASTHobbs2014}
{Hobbs}, G., {Dai}, S., {Manchester}, R.~N., {et~al.} 2014, ArXiv e-prints

\bibitem[{{Hu} {et~al.}(2013){Hu}, {Nan}, {Zhu}, \& {Li}}]{FASTHu2013}
{Hu}, J., {Nan}, R., {Zhu}, L., \& {Li}, X. 2013, Measurement Science and
  Technology, 24, 095006

\bibitem[{{Jiang} {et~al.}(2015){Jiang}, {Nan}, {Qian}, \&
  {Yue}}]{FASTJiang2015}
{Jiang}, P., {Nan}, R.-D., {Qian}, L., \& {Yue}, Y.-L. 2015, ArXiv e-prints

\bibitem[{{Kogut}(2012)}]{Kogut:2012}
{Kogut}, A. 2012, \apj, 753, 110

\bibitem[{{Lawson} {et~al.}(1987){Lawson}, {Mayer}, {Osborne}, \&
  {Parkinson}}]{Lawson:1987}
{Lawson}, K.~D., {Mayer}, C.~J., {Osborne}, J.~L., \& {Parkinson}, M.~L. 1987,
  \mnras, 225, 307

\bibitem[{{Lewis} {et~al.}(2000){Lewis}, {Challinor}, \& {Lasenby}}]{CAMB}
{Lewis}, A., {Challinor}, A., \& {Lasenby}, A. 2000, \apj, 538, 473

\bibitem[{{Liddle}(1999)}]{Liddle:1999}
{Liddle}, A.~R. 1999, in High Energy Physics and Cosmology, 1998 Summer School,
  ed. A.~{Masiero}, G.~{Senjanovic}, \& A.~{Smirnov}, 260

\bibitem[{Morrison {et~al.}(1979)Morrison, Billingham, \& Wolfe}]{SETI:2007}
Morrison, P., Billingham, J., \& Wolfe, J., eds. 1979, The Search for
  Extraterrestrial Intelligence, NASA SP-419 (Dover: New York)

\bibitem[{{Nan} \& {Li}(2013)}]{FASTNan2013}
{Nan}, R. \& {Li}, D. 2013, Materials Science and Engineering Conference
  Series, 44, 012022

\bibitem[{Nan {et~al.}(2011)Nan, Li, Jin, Wang, Zhu, Zhu, Zhang, Yue, \&
  Qian}]{Nan:2011}
Nan, R., Li, D., Jin, C., {et~al.} 2011, Int. J. Modern Phys. D, 20, 989

\bibitem[{{Pacholczyk}(1970)}]{Pacholczyk:1970}
{Pacholczyk}, A.~G. 1970, {Radio astrophysics. Nonthermal processes in galactic
  and extragalactic sources} (Freeman)

\bibitem[{Peng {et~al.}(2009)Peng, Jin, Wang, Zhu, Zhu, Zhang, \&
  Nan}]{FAST_prep}
Peng, B., Jin, C., Wang, Q., {et~al.} 2009, Proceedings of the IEEE, 97, 1391

\bibitem[{{Peterson} {et~al.}(2009){Peterson}, {Aleksan}, {Ansari}, {Bandura},
  {Bond}, {Bunton}, {Carlson}, {Chang}, {DeJongh}, {Dobbs}, {Dodelson},
  {Darhmaoui}, {Gnedin}, {Halpern}, {Hogan}, {Le Goff}, {Liu}, {Legrouri},
  {Loeb}, {Loudiyi}, {Magneville}, {Marriner}, {McGinnis}, {McWilliams},
  {Moniez}, {Palanque-Delabruille}, {Pasquinelli}, {Pen}, {Rich}, {Scarpine},
  {Seo}, {Sigurdson}, {Seljak}, {Stebbins}, {Steffen}, {Stoughton}, {Timbie},
  {Vallinotto}, \& {Teche}}]{Peterson:2009}
{Peterson}, J.~B., {Aleksan}, R., {Ansari}, R., {et~al.} 2009, in Astronomy,
  Vol. 2010, astro2010: The Astronomy and Astrophysics Decadal Survey, 234

\bibitem[{{Planck Collaboration}(2014)}]{Planck-Collaboration:2013aa}
{Planck Collaboration}. 2014, \aap, 571, A16

\bibitem[{{Planck Collaboration}(2015)}]{Planck-Collaboration2015aa}
{Planck Collaboration}. 2015, preprint (arXiv:1502.01589)

\bibitem[{{Platania} {et~al.}(2003){Platania}, {Burigana}, {Maino}, {Caserini},
  {Bersanelli}, {Cappellini}, \& {Mennella}}]{Platania:2003}
{Platania}, P., {Burigana}, C., {Maino}, D., {et~al.} 2003, \aap, 410, 847

\bibitem[{{Reich} \& {Reich}(1988)}]{Reich:1988}
{Reich}, P. \& {Reich}, W. 1988, \aaps, 74, 7

\bibitem[{{Sahni} \& {Starobinsky}(2000)}]{Sahni:2000}
{Sahni}, V. \& {Starobinsky}, A. 2000, Int. J. Modern Phys. D, 9, 373

\bibitem[{{Santos} {et~al.}(2015){Santos}, {Bull}, {Alonso}, {Camera},
  {Ferreira}, {Bernardi}, {Maartens}, {Viel}, {Villaescusa-Navarro}, {Abdalla},
  {Jarvis}, {Metcalf}, {Pourtsidou}, \& {Wolz}}]{Santos2015}
{Santos}, M., {Bull}, P., {Alonso}, D., {et~al.} 2015, Advancing Astrophysics
  with the Square Kilometre Array (AASKA14), 19

\bibitem[{{Seiffert} {et~al.}(2011){Seiffert}, {Fixsen}, {Kogut}, {Levin},
  {Limon}, {Lubin}, {Mirel}, {Singal}, {Villela}, {Wollack}, \&
  {Wuensche}}]{ARCADE2skybrightness}
{Seiffert}, M., {Fixsen}, D.~J., {Kogut}, A., {et~al.} 2011, \apj, 734, 6

\bibitem[{{Seo} {et~al.}(2010){Seo}, {Dodelson}, {Marriner}, {Mcginnis},
  {Stebbins}, {Stoughton}, \& {Vallinotto}}]{Seo:2010}
{Seo}, H.-J., {Dodelson}, S., {Marriner}, J., {et~al.} 2010, \apj, 721, 164

\bibitem[{{Shaver} {et~al.}(1999){Shaver}, {Windhorst}, {Madau}, \& {de
  Bruyn}}]{Shaver:1999}
{Shaver}, P.~A., {Windhorst}, R.~A., {Madau}, P., \& {de Bruyn}, A.~G. 1999,
  \aap, 345, 380

\bibitem[{{Shaw} {et~al.}(2014){Shaw}, {Sigurdson}, {Pen}, {Stebbins}, \&
  {Sitwell}}]{Shaw:2014}
{Shaw}, J.~R., {Sigurdson}, K., {Pen}, U.-L., {Stebbins}, A., \& {Sitwell}, M.
  2014, \apj, 781, 57

\bibitem[{{Smoot} {et~al.}(1992){Smoot}, {Bennett}, {Kogut}, {Wright}, {Aymon},
  {Boggess}, {Cheng}, {de Amici}, {Gulkis}, {Hauser}, {Hinshaw}, {Jackson},
  {Janssen}, {Kaita}, {Kelsall}, {Keegstra}, {Lineweaver}, {Loewenstein},
  {Lubin}, {Mather}, {Meyer}, {Moseley}, {Murdock}, {Rokke}, {Silverberg},
  {Tenorio}, {Weiss}, \& {Wilkinson}}]{COBE:1992}
{Smoot}, G.~F., {Bennett}, C.~L., {Kogut}, A., {et~al.} 1992, \apjl, 396, L1

\bibitem[{{Spergel} {et~al.}(2003){Spergel}, {Verde}, {Peiris}, {Komatsu},
  {Nolta}, {Bennett}, {Halpern}, {Hinshaw}, {Jarosik}, {Kogut}, {Limon},
  {Meyer}, {Page}, {Tucker}, {Weiland}, {Wollack}, \& {Wright}}]{WMAP:2003}
{Spergel}, D.~N., {Verde}, L., {Peiris}, H.~V., {et~al.} 2003, \apjs, 148, 175

\bibitem[{{Tegmark} {et~al.}(2000){Tegmark}, {Eisenstein}, {Hu}, \& {de
  Oliveira-Costa}}]{Tegmark:2000}
{Tegmark}, M., {Eisenstein}, D.~J., {Hu}, W., \& {de Oliveira-Costa}, A. 2000,
  \apj, 530, 133

\bibitem[{{Wang} {et~al.}(2006){Wang}, {Tegmark}, {Santos}, \&
  {Knox}}]{Wang:2006}
{Wang}, X., {Tegmark}, M., {Santos}, M.~G., \& {Knox}, L. 2006, \apj, 650, 529

\bibitem[{{Willis} {et~al.}(1977){Willis}, {Oosterbaan}, {Le Poole}, {de
  Ruiter}, {Strom}, {Valentijn}, {Katgert}, \&
  {Katgert-Merkelijn}}]{Willis:1977}
{Willis}, A.~G., {Oosterbaan}, C.~E., {Le Poole}, R.~S., {et~al.} 1977, in IAU
  Symposium, Vol.~74, Radio Astronomy and Cosmology, ed. D.~L. {Jauncey}, 39

\bibitem[{{Wolleben} {et~al.}(2009){Wolleben}, {Landecker}, {Carretti},
  {Dickey}, {Fletcher}, {Gaensler}, {Han}, {Haverkorn}, {Leahy},
  {McClure-Griffiths}, {McConnell}, {Reich}, \& {Taylor}}]{Wolleben:2009}
{Wolleben}, M., {Landecker}, T.~L., {Carretti}, E., {et~al.} 2009, in IAU
  Symposium, Vol. 259, IAU Symposium, ed. K.~G. {Strassmeier}, A.~G.
  {Kosovichev}, \& J.~E. {Beckman}, 89--90

\end{thebibliography}

\end{document}